\documentclass[pdflatex,sn-mathphys-ay]{sn-jnl}

\usepackage{graphicx}%
\usepackage{multirow}%
\usepackage{amsmath,amssymb,amsfonts}%
\usepackage{amsthm}%
\usepackage{mathrsfs}%
\usepackage[title]{appendix}%
\usepackage{xcolor}%
\usepackage{textcomp}%
\usepackage{manyfoot}%
\usepackage{booktabs}%
\usepackage{algorithm}%
\usepackage{algorithmicx}%
\usepackage{algpseudocode}%
\usepackage{listings}%
\usepackage[normalem]{ulem}

\theoremstyle{thmstyleone}%
\theoremstyle{thmstyletwo}%

\theoremstyle{thmstylethree}%

\begin{document}
\title[MHD waves with mixed properties ]{MHD waves with mixed properties/Alfv\'{e}n waves with pressure variations: a review}


\author*[1]{\fnm{Marcel} \sur{Goossens}}
\equalcont{All the authors contributed equally to this work.}

\author*[2,3]{\fnm{I\~nigo} \sur{Arregui}}\email{iarregui@iac.es}
\equalcont{All the authors contributed equally to this work.}

\author[4,5]{\fnm{Roberto} \sur{Soler}}
\equalcont{All the authors contributed equally to this work.}

\author[4,5]{\fnm{Jaume} \sur{Terradas}}
\equalcont{All the authors contributed equally to this work.}

\author[1]{\fnm{Tom} \sur{Van Doorsselaere}}
\equalcont{All the authors contributed equally to this work.}

\affil[1]{\orgdiv{Centre  for mathematical Plasma Astrophysics, Department of Mathematics}, \orgname{KU Leuven}, \orgaddress{\street{Celestijnenlaan 200B Bus 2400}, \city{Leuven}, \postcode{B-3001}, \country{Belgium}}}

\affil[2]{\orgdiv{Instituto de Astrof\'{\i}sica de Canarias}, \orgaddress{\street{C/ V\'{\i}a L\'actea s/n}, \city{La Laguna}, \postcode{E-38205}, \state{Tenerife}, \country{Spain}}}

\affil[3]{\orgdiv{Departamento de Astrof\'{\i}sica},  \orgname{Universidad de La Laguna}, \orgaddress{\city{La Laguna}, \postcode{E-38206}, \state{Tenerife}, \country{Spain}}}

\affil[4]{\orgdiv{Departament de F\'{\i}sica}, \orgname{Universitat de les Illes Balears},  \orgaddress{\city{Palma de Mallorca}, \postcode{E-07122}, \country{Spain}}}

\affil[5]{\orgdiv{Institute of Applied Computing \& Community Code (IAC3)},  \orgaddress{\city{Palma de Mallorca}, \postcode{E-07122}, \country{Spain}}}


\abstract{Non-uniformity plays an important role for MHD waves.  For a uniform plasma of infinite extent the MHD waves can be subdivided in two classes with distinct properties. The first class contains the Alfv\'{e}n waves. The Alfv\'{e}n waves are incompressible and propagate parallel vorticity. They do not have a parallel component of displacement,  they do not cause variations in pressure  and are driven by magnetic tension only. The second class contains the magneto-sonic waves. They  are compressible and have a parallel component of displacement. They do not propagate parallel vorticity and  are driven  by pressure and magnetic tension.  In non-uniform plasmas the situation can be very different. The clear division between Alfv\'{e}n waves and magneto-sonic waves is no longer present.  In a given part of the equilibrium an MHD wave can strongly resemble a magneto-sonic wave  with little or no resemblance to  Alfv\'{e}n waves;  while in another part of the equilibrium the MHD wave is practically an Alfv\'{e}n wave, which has the amazing property of being accompanied by variations in pressure. } 

\keywords{Magnetohydrodynamics (MHD), Sun: atmosphere, Sun: magnetic fields, Sun: corona, Sun:
oscillations, waves}

\maketitle

\section{Introduction}\label{sec:intro}
The concept of MHD waves with mixed properties does not seem to be well known in the Solar Plasma Physics community. This became again clear during the Solar Physics sessions  of the 2024-AAPPS-DPP2024 conference. Even when we take into account the observation by \cite{cally24} that mode conversion can be an alternative interpretation, mode conversion from e.g. fast to Alfv\'en actually is a wave with mixed properties that changes from a wave with dominant fast properties to a wave with dominant  Alfv\'en properties.  The properties under scrutiny are  borrowed from the analysis of MHD waves for uniform plasmas of infinite extent. For a uniform plasma of infinite extent the MHD waves can be put in separate boxes as magneto-sonic waves and Alfv\'{e}n waves. Apparently this clear division is much favoured  when MHD waves are studied; even when the MHD waves live in non-uniform plasmas. 

The fact that MHD waves with mixed properties are not well known in the solar plasma physics community is rather surprising since they were already present in early studies on resonant absorption of Alfv\'{e}n waves. This can be clearly seen in the numerical study of \cite{poedts89}. In their Figures 3-4-5, the total pressure and the three components of the Lagrangian displacement are non-zero.  For a straight field the mixed properties are due to the perturbation of total pressure. This result can be deduced from the remark by \cite{hasegawa82} in their book ``The Alfv\'{e}n wave". The basic characteristics of the ideal Alfv\'{e}n wave is that total pressure in the fluid remains constant during the passage of the wave as a consequence of the incompressibility condition. For an inhomogeneous medium, however, the total pressure couples with the dynamics of the motion, and the assumption of neglect of pressure perturbations becomes invalid. However, this does not mean that the MHD wave is a classic magneto-sonic wave. So where do the mixed properties come from? The physical reason is stratification transverse to the equilibrium magnetic field. From a mathematical point of view the mixed properties arise because the equations for the different wave variables are coupled.

For recent reviews on MHD waves in the structured solar atmosphere we refer to \cite{magyar24} and \cite{soler24}. In this review, we emphasize the changes that appear when we move from uniform to non-uniform plasmas.  We start with the analysis of MHD waves in a uniform plasma of infinite extent and summarise the properties of relevant wave quantities in Section~\ref{sec:2}. Particular attention is given to the components of displacement, vorticity, pressure  and compression.  In Section~\ref{sec:3} we shall turn to MHD waves in non-uniform plasmas  in a static equilibrium. The non-uniform equilibrium is a straight 1-dimensional plasma column. The phenomenon of mixed properties is discussed in this section. Section~\ref{sec:4} discusses resonant absorption as a clear example of MHD waves with mixed properties.

\section{Linear MHD waves of a uniform plasma of infinite extent}\label{sec:2}

Non-uniformity plays an important role for MHD waves.  In order to highlight  this role  we give a short revision of MHD waves in a uniform plasma of infinite extent.  For a uniform plasma of infinite extent the MHD waves can be subdivided in two  classes with distinct properties. The first class contains the Alfv\'{e}n waves. The Alfv\'{e}n waves are incompressible and propagate parallel vorticity. They do not have a parallel component of displacement and no pressure perturbations and are driven by magnetic tension only,  The second class contains the magneto-sonic waves. They  are compressible and have a parallel component of displacement. They do not propagate parallel vorticity and  are driven  by pressure and magnetic tension.  MHD waves in a uniform plasma of infinite extent  are discussed in many textbooks \citep[see e.g.][]{goedbloed83,goossens03,walker04,goedbloed04}.  In non-uniform plasmas the situation can be very different. The clear division between Alfv\'{e}n waves and magneto-sonic waves is no longer present.

In a uniform medium the  equilibrium quantities are constant. The  constant magnetic field 
\begin{equation}
\vec{B_0} = B_0 \vec{1}_z,
\label{B0}
\end{equation}
is used to define the direction of the $z$-axis of a Cartesian system of coordinates. The equilibrium density and pressure are also constant  
\begin{equation}
p_0=  \mbox{constant}, \; \rho_0=\mbox{constant}.
\label{p0rho0}
\end{equation}

In what follows $\vec{\xi}$ is the Lagrangian displacement.  In the present subsection the background is static and uniform. As a consequence solutions can be obtained in the form of plane-harmonic waves and  $\vec{\xi}$ is written 
as 
\begin{equation}
\vec{\xi} (\vec{r};t) =   \vec{\hat{\xi}} \exp(i(\vec{k}\cdot\vec{r} - \omega t))
 =\vec{\hat{\xi}} \exp(i(k_x x + k_y y + k_z z - \omega t)).
              \label{PlaneWaves}         
\end{equation}
Here  $\vec{\hat{\xi}}$ is the constant amplitude of $\vec{\xi}$, $\vec{k} = k_x \vec{1}_x + k_y \vec{1}_y + k_z \vec{1}_z$ is  the wave vector, and   $\omega$  is  the frequency of the wave.  In what follows the hat on $\vec{\xi}$ will be dropped.  Since the constant magnetic field defines a preferred direction a clever choice of dependent wave  variables is  $\xi_z, Y, Z$ defined as

\begin{eqnarray}
\xi_z & = & \mbox{component of $\vec{\xi}$ parallel to } \vec{B}_0, \nonumber \\
\nabla \cdot \vec{\xi} = i\; \vec{k} \cdot \vec{\xi} = i \; Y& = &  \mbox{compression}, \nonumber \\
(\nabla \times \vec{\xi} )_z = i \;(\vec{k} \times \vec{\xi})_z  = i  \;Z & =& \mbox{component of vorticity parallel to} \;\vec{B}_0. 
\label{XYZWaves}     
\end{eqnarray}

In terms of these variables the equations for linear ideal MHD waves can be written as \citep{goossens12}
\begin{eqnarray}
\omega^2 \xi_z - k_z v_{S}^2  Y&  = &  0,  \nonumber \\
k^2 v_A^2 k_z  \xi_z + (\omega^2 -  k^2 (v_A^2 + v_{S}^2)) Y&  = &  0, \nonumber\\
(\omega^2 - \omega_{ A}^2) Z &  = &  0.
\label{EqXYZ}     
\end{eqnarray}
Here $v_A$, $v_S$ are the Alfv\'{e}n velocity and the velocity of sound, respectively. They are defined as 
\begin{equation}
 v_A^2 = \frac{\displaystyle B_0^2}{\displaystyle \mu \; \rho_0}, \;\;
v_S^2 = \frac{\displaystyle \gamma p_0}{\rho_0}.
\label{VAVS}
\end{equation}
 $\omega_A$  is the local Alfv\'{e}n frequency, it is defined as 
\begin{equation}
 \omega_A^2  = \frac{\displaystyle (\vec{k} \cdot
\vec{B})^2}{\displaystyle \mu \rho} \;=\;k_z^2\;v_A^2.
\label{AlfvenFre1}
\end{equation}
In a uniform plasma $v_A, v_S, \omega_A$ are constant. In a non-uniform plasma these quantities depend on position. 

\subsection{Classic Alfv\'{e}n waves}\label{aws}

The system~(\ref{EqXYZ}) consists of two  uncoupled subsets of equations.  The first  subset is the third equation for the variable $Z$. The second  subset  contains the wave variables $\xi_z$ and $Y$.  The first type of MHD waves are characterized by 
\begin{equation}
 Y= 0, \; Z\neq 0, \; \xi_z = 0,\;\; \omega^2 = \omega_A^2.
\label{AlfvenW1}
\end{equation}
They are the classic Alfv\'{e}n waves. The eigenfrequencies associated with the Alfv\'{e}n waves~(\ref{AlfvenW1}) are infinitely degenerate as they only depend on the parallel component of the wave vector $\vec{k}$. Alfv\'{e}n waves   do not cause compression and have no component of the displacement parallel to the magnetic field. They are the only waves that propagate parallel vorticity in a uniform plasma of infinite extent.  The only restoring force is the magnetic tension force. Note also that Alfv\'{e}n waves in a uniform plasma of infinite extent exist for any wave vector  $\vec{k} = (k_x, k_y, k_z)$. 

The only exception to the condition that Alfv\'{e}n waves have non-zero parallel vorticity occurs for waves with $k_x =0, \;k_y =0$ and $k_z \neq 0$. Obviously $Z =0$.  The solutions in this particular case are:

$\bullet$ The Alfv\'{e}n waves 
\begin{equation}
Y =0, \;Z=0, \;\xi_z =0, \;\omega^2 =  \omega_A^2 , \; \vec{\xi} = \xi_x \vec{1}_x + \xi_y \vec{1}_y.
\label{AlfExc}
\end{equation}

$\bullet$ The magneto-sonic  waves 
\begin{equation}
Y \neq 0, \;Z=0, \;\xi_z \neq 0, \;\omega^2 =  k_z ^2 v_S^2, \;, \; \vec{\xi} = \xi_z \vec{1}_z.
\label{AlfExc}
\end{equation}

Let us now consider the case that $k_x$ and $k_y$ are not simultaneously zero.  The displacement $\vec{\xi}$ for Alfv\'{e}n waves is
\begin{equation}
\vec{\xi}_A  =( - \frac{k_y}{k_x} \vec{1}_x + \vec{1}_y) \xi_y \;=
\;(\vec{1}_x    - \frac{k_x}{k_y} \vec{1}_y) \xi_x.
\label{DisplacementAlfven}
\end{equation}

For $k_y =0$ we obtain the popular result $\vec{\xi}_A  = \xi_y \vec{1}_y$.
These $y$-independent  Alfv\'{e}n waves are a special case.  The case $k_y = 0$ in Cartesian geometry corresponds to axisymmetric waves with $m=0$ in cylindrical geometry. The case $k_y \neq 0$ in Cartesian geometry corresponds to non-axisymmetric waves with $m \neq 0$ in cylindrical geometry, with $m$ the azimuthal wave number.  For a wave vector with both horizontal  components different from zero both horizontal components of  the displacement vector   are non-zero. Let us now keep   $k_y \neq 0, \; k_z \neq 0$ and finite and mimic a situation with non-uniformity in the $x$-direction and a resonant condition where $\lim k_x \rightarrow + \infty$ so that $\mid k_y \mid \ll \mid k_x \mid, \mid k_z  \mid \ll  \mid k_x  \mid $.   Find  then 
\begin{equation}
\frac{\mid \xi_y \mid}{\mid \xi_x \mid} = \frac{\mid k_x \mid}{\mid k_y \mid} \gg 1,  \;\;\vec{\xi}_A \approx \xi_y\;\vec{1}_y 
\label{DisplacementAlfven2}.
\end{equation} 
The motion in this  Alfv\'{e}n wave is predominantly in the $y$-direction and rapidly varying in the $x$-direction.  The displacement~(\ref{DisplacementAlfven2}) is  $y$-dependent  because of the factor $\exp(i k_y y)$ with $k_y \neq 0$.  The $\approx$ sign means that the two components $(\xi_x, \xi_y)$ are non-zero  but $\xi_y$ is far larger in absolute value than $\xi_x$. The two components  are needed to satisfy the incompressibility condition.

For a general wave vector $\vec{k} = (k_x, k_y,k_z)^t$  the three components of vorticity $\nabla \times \vec{\xi}$ are non-zero. In addition to the parallel component  $(\nabla \times \vec{\xi})_z $ also the components in planes normal to $\vec{B}_0 $ are non-zero:
\begin{equation}
(\nabla \times \vec{\xi})_z = i (k_x \xi_y - k_y \xi_x), \;\;(\nabla  \times \vec{\xi})_x = - i \;k_z  \xi_y, \;\; (\nabla \times \vec{\xi})_y =  i \;k_z  \xi_x, \;\; \xi_x = - \frac{k_y}{k_x} \xi_y.
\label{VorticityAlfven}
\end{equation}

For our later discussion on resonant Alfv\'{e}n waves it is instructive to look at the components of vorticity $\nabla \times \vec{\xi}$ under conditions that mimic resonant behaviour, i.e. when 
$\mid k_y\mid \ll \mid k_x\mid, \mid k_z\mid \ll  \mid k_x\mid$  and find that

\begin{equation}
\nabla \times \vec{\xi} \approx (\nabla \times \vec{\xi} )_z \;\vec{1}_z \approx i k_x \xi_y \;\vec{1}_z
\label{VorticityAlfven2}.
\end{equation}
Here also the $\approx$ sign means that the three components $(\nabla \times \vec{\xi} )$ are non-zero  but the parallel component  is far larger in absolute value than the two horizontal components.   Equations~(\ref{DisplacementAlfven2}) and (\ref{VorticityAlfven2}) will come back in our discussion on resonant Alfv\'{e}n waves in Section~\ref{global}.

The flow of energy is exclusively along the magnetic field lines with velocity $v_A$. The group velocity of Alfv\'{e}n waves is 
\begin{equation}
\vec{v}_{gr,A} = v_A \vec{1}_z.
\label{GrVAlfven}
\end{equation}

\subsection{Slow and fast magneto-sonic waves}

The second class of MHD waves corresponds to 
\begin{equation}
Y \neq 0, \; Z= 0, \; \xi_z = \xi_{\parallel} \neq 0.
\label{FastSlowW1}
\end{equation}
They are the magneto-sonic waves. They cause compression (and pressure perturbations) but do not propagate  parallel vorticity. 
However, they cause  horizontal vorticity. Their displacement  has a component parallel to  the magnetic field that is driven by the magnetic pressure force.  The dispersion relation is 

\begin{equation}
(\omega^2)^2 - k^2 (v_S^2 + v_A^2) \omega^2 + k_z^2\;k^2\;v_S^2\;v_A^2 = 0.
\label{DRFastSlow}
\end{equation}
The solutions for the eigenfrequencies are 
\begin{equation}
\omega^2 = \omega^2_{sl,f} = \frac{\displaystyle k^2 (v_S^2 + v_A^2)}
{\displaystyle 2} \left \{ 1 \pm \left (1 - \frac{\displaystyle 4 \omega_C^2}
{\displaystyle k^2 (v_S^2 + v_A^2)} \right )^{1/2} \right \}.
\label{FastSlowFre1}
\end{equation}
$k^2 = k_x^2 + k_y^2 + k_z^2$ , $\omega_C $ and $v_C  $ are the cusp frequency, and the cusp velocity 
\begin{equation}
\omega_C^2 = \frac{\displaystyle v_S^2}{\displaystyle v_S^2 + v_A^2}
\omega_A^2  = k_{\mid \mid}^2 v_C^2  =  k_z^2 v_C^2, \;\;
v_C^2 = \frac{\displaystyle v_S^2 v_A^2}{\displaystyle v_S^2 +
v_A^2}.
\label{CuspFre1}
\end{equation}
In Equation~(\ref{FastSlowFre1}) ``sl" corresponds to  the minus sign,  and ``f" to  the plus sign. The corresponding waves are the slow and fast magneto-sonic waves.  The frequencies of the magneto-sonic waves depend on the three components $(k_x, k_y, k_z)$ of the wave vector $\vec{k}$. They  depend in the same way  on $k_x$ and $k_y$  because of  isotropy in the planes normal to $\vec{B}_0$.  It is instructive to consider the variation of  $\omega^2_{sl,f}$ as function of $k_x$ for fixed values of $(k_y, k_z)$. The cut-off frequencies $\omega_{I}, \omega_{II}$ are defined as 
\begin{equation}
\omega_{I}^2 = \omega_{sl}^2(k_x =0, k_y, k_z), \;\;\omega_{II}^2 = \omega_{f}^2(k_x =0, k_y, k_z).
\label{Cutoff1}
\end{equation}
Also 
\begin{equation}
\mbox{lim}_{k_x \rightarrow \infty} \omega_{sl}^2 = \omega_C^2, \;\; \mbox{lim}_{k_x  \rightarrow \infty} \omega_{f}^2 = \infty.
\label{Limitkxinfinity}
\end{equation}

The cut-off frequencies $\omega_{I}, \;\omega_{II}$ and the characteristic  frequencies $\omega_A,\; \omega_C$  define  a sequence of  inequalities 
\begin{equation}
\omega_C^2\leq \omega_{sl}^2 \leq \omega_{I}^2 \leq \omega_A^2 \leq \omega_{II}^2 \leq \omega_{f }^2 < + \infty.
\label{3Subspectra}
\end{equation}
Hence the spectrum of linear motions of a uniform plasma of infinite extent can be divided  in a slow subspectrum 
$]\omega_C, \;\omega_{I}]$, a degenerate  Alfv\'{e}n  point spectrum $\omega_A$ and a fast subspectrum $[\omega_{II}, \;\;+ \infty[$.  The first equality in~(\ref{Limitkxinfinity})  means that $\omega_C$ is an accumulation point of the slow subspectrum.

The magneto-sonic waves  are driven by tension and pressure forces and cause variations in density, pressure and horizontal vorticity. The solutions for the eigenfunctions  are 
\begin{equation}
\vec{\xi}_{sl,f} = \left( \vec{1}_x + \frac{k_y}{k_x} \;\vec{1}_y + \frac{ \omega^2_{sl,f} - k^2 v_A^2}{\omega^2_{sl,f}} \; \frac{k_z}{k_x} 
\vec{1}_z\right) \; \xi_x
\label{XiFastSlowF1}
\end{equation}
or equivalently 
\begin{equation}
\vec{\xi}_{sl,f} = \left(\frac{\omega^2_{sl,f} }{\omega^2_{sl,f} - k^2 v_A^2}\;\frac{k_x}{k_z} \vec{1}_x + \frac{\omega^2_{sl,f} }{\omega^2_{sl,f} - k^2 v_A^2}\;\frac{k_y}{k_z} \vec{1}_y +\vec{1}_z \right)\xi_z.
\label{XiFastSlowF2}
\end{equation}
The popular view is that the horizontal motion $(\xi_x, \xi_y)$ is the dominant motion for fast waves while  the parallel 
motion $\xi_z$ is the dominant motion for slow waves.  In order to point  out that this is not the general rule,  $\xi_x$ is used  as the measuring unit in~(\ref{XiFastSlowF1}) and  $\xi_z$ in~(\ref{XiFastSlowF2}). It is straightforward to show that  in general the parallel component in~(\ref{XiFastSlowF1}) is not small compared to the horizontal components, and  similarly  that  the horizontal components in~(\ref{XiFastSlowF2}) are not per se much smaller than the parallel component.  However, for  strong magnetic fields, i.e. $v_A \gg v_S$ it can be shown that
\begin{equation}
\vec{\xi}_{f} \approx  \left( \vec{1}_x + \frac{k_y}{k_x} \;\vec{1}_y \right) \; \xi_x; \;\;
\vec{\xi}_{sl}\approx   \xi_z \vec{1}_z. 
\label{XiFastSlowF3}
\end{equation}
The popular view corresponds  to the limiting case of a strong field.

The parallel component of vorticity $(\nabla \times  \vec{\xi})_z = i Z $ is of course zero. However the horizontal components are non-zero
\begin{equation}
\nabla \times \vec{\xi} = - i\; k_z \;\frac{k^2 \;v_A^2}{\omega^2_{sl,f}}\;\xi_x\; \left(\frac{k_y}{k_x} \vec{1}_x  - \vec{1}_y\right).
\label{VorticityFastSlowF}
\end{equation}

For  $k_y =0$ the expressions (\ref{XiFastSlowF1}) for the displacement $\vec{\xi}$ and (\ref{VorticityFastSlowF}) for vorticity $\nabla \times \vec{\xi}$  can be simplified to 

\begin{equation}
\vec{\xi}_{sl,f} = \left( \vec{1}_x + \frac{ \omega^2_{sl,f} - k^2 v_A^2}{\omega^2_{sl,f}} \; \frac{k_z}{k_x}  \vec{1}_z\right) \; \xi_x, \;\;
\nabla \times \vec{\xi} = i\; k_z \;\frac{k^2 \;v_A^2}{\omega^2_{sl,f}}\;\xi_x\; \vec{1}_y.
\label{XiVorticity+FastSlow}
\end{equation}
Keep  $k_y \neq 0, \; k_z \neq 0$ and finite and mimic a situation with non-uniformity in the $x$-direction and a turning point where $k_x  = 0$ and  find  
\begin{equation}
\vec{\xi}_{sl,f} = \left( \frac{\omega^2_{I,II} }{\omega^2_{I,II} - k^2 v_A^2}\;\frac{k_y}{k_z} \vec{1}_y +\vec{1}_z \right)\xi_z.
\label{XiFastSlowF3}
\end{equation}

The plasma pressure and magnetic pressure variations are in phase for the fast waves and in anti-phase for the slow waves. Propagation of phase and energy is fairly isotropic for the fast waves. There are relatively  small variations in the absolute values  of the  phase speeds and the group speeds as function of the direction of propagation vector $\vec{k}$ and there are small differences between  the directions of the phase speed and group speed. Matters are different for the slow waves. In particular
\begin{equation}
\lim_{k_x \rightarrow \infty} \vec{v}_{gr, sl} = v_C \vec{1}_z.
\end{equation}
This result tells us that propagation of phase perpendicular to the magnetic field  is associated with flow of energy along the magnetic field. In general, flow of energy is highly anisotropic for slow waves and confined to a narrow cone around the magnetic field lines.  

In summary for a uniform plasma of infinite extent the division is clear.  The waves have either  parallel  vorticity  and no compression (i.e. variation of pressure)  and no parallel displacement, these are the Alfv\'{e}n waves;  or they have compression (i.e. pressure perturbations)  and parallel displacement and no parallel vorticity, they are  magneto-sonic waves. There are no waves with compression and  parallel vorticity at the same time. There is no mixing of properties.  A comment on compression is in order here. For a plasma with constant density  zero/non-zero compression implies zero/non-zero pressure perturbation. However, this is not the case for MHD waves on an equilibrium with non-constant equilibrium density since there incompressible motions can have non-zero pressure perturbations.

\subsection{MHD waves for a pressureless plasma}

To facilitate seamless access, we present here the equations for the particular case of a pressureless plasma with $v_S^2 = 0$. The solutions for the magneto-sonic waves are 

\begin{eqnarray}
\omega^2_{C} & = & 0,\;\;\omega^2_{sl} = 0, \;\;\omega^2_{f} = k^2 v_A^2, \;\; \nonumber \\
\xi_z &  = & 0, \;\;\vec{\xi}_{f} =  \left( \vec{1}_x + \frac{k_y}{k_x} \;\vec{1}_y \right) \; \xi_x, \;\;\;
\nabla \times \vec{\xi} = - i\; k_z \;\;\xi_x\; \left(\frac{k_y}{k_x} \vec{1}_x  - \vec{1}_y\right).
\label{XiFastSlowF4}
\end{eqnarray}
In this situation there are no slow waves and the fast magneto-sonic waves  have no parallel motions. The parallel motions are driven by the gradient of plasma pressure and here plasma pressure vanishes by assumption.  The absence of slow waves and of parallel motions is  a general result for a pressureless plasma. 
In what follows no particular attention will be devoted to pressureless plasmas. The equations for MHD waves for a pressureless plasma are easily obtained by putting $v_S^2 =0$ in  the general equations.

\section{Mixed properties in non-uniform plasmas}\label{sec:3}

\subsection{General case for a twisted equilibrium magnetic field}

Non-uniformity causes decisive changes in the properties of MHD waves compared to the properties in uniform plasmas that we discussed in the previous section. The MHD waves have now  mixed properties.  Mixed properties mean that the general rule is that MHD waves propagate both parallel vorticity as in classic Alfv\'{e}n waves and pressure perturbations as in classic magneto-sonic waves. Also  in a non-uniform plasma  the MHD waves behave differently in different parts of the plasma. In addition the MHD waves can undergo resonant behaviour.  The mixed properties arise because in an inhomogeneous plasma the  Eulerian perturbation of total pressure couples with the dynamics of the motion \citep{hasegawa82}. Mathematically this is reflected by the fact  that the differential equations for the radial component of the Lagrangian displacement $\xi_r$  and the Eulerian perturbation of total pressure $P'$  are coupled to algebraic equations for compression $\nabla \cdot \vec{\xi}$, the parallel and perpendicular projections of the Lagrangian displacement $\xi_{\parallel}, \xi_{\perp}$,  and vorticity  $\nabla \times \vec{\xi}$. The coupling of the equations is due to  the coupling functions $C_A$ and $C_S$ which were introduced by \cite{sakurai91} in their study of resonant absorption. The relevance of the coupling functions goes beyond resonant absorption. The spatial behaviour of the coupling functions and of the  local Alfv\'{e}n frequency  $\omega_A$  and local cusp frequency $\omega_C$ determine the spatial behaviour of  the various components of velocity and  vorticity and of compression.  The simultaneous presence of compression and vorticity is hard to avoid. 

The phenomenon  of mixed properties follows from the fact that the equations that describe the linear motions are coupled, unlike for the case of a uniform plasma of infinite extent.   In particular the focus is on MHD  waves on 1-D cylindrical plasma columns. The equilibrium model is a  straight cylindrical plasma column of radius $R$ in static equilibrium. In what follows we use  cylindrical
coordinates ($r, \varphi, z$).  The magnetic field has both an axial and an azimuthal component
\begin{equation}
\vec{B_0} = B_{z,0} \vec{1}_z  + B_{\varphi,0} \vec{1}_{\varphi}.
\label{B0}
\end{equation}
The equilibrium density $\rho_0(r)$, equilibrium pressure $p_0(r)$ and the components of the equilibrium magnetic field $B_{z,0}(r), B_{\varphi,0}(r)$ are functions of $r$ or constant. The equilibrium quantities satisfy the equation of static equilibrium
\begin{equation}
 \frac{d}{dr} (p_0 +\frac{B_0^{2}}{2\mu})
=-\frac{B_{\varphi,0}^2}{\mu r}, \;\;\; B_0^2  = B_{\varphi,0}^2 +B_{z,0}^2.
\label{ForceBalance}
\end{equation}
In a nonuniform plasma $v_S^2$, $v_A^2$, $\omega_A^2$, and $\omega_C^2$ are functions of position. 
In what follows $P' =  p' +  \vec{B_0}\cdot \vec{B'}/ \mu $ is the  Eulerian perturbation of total pressure;  $p'$ is the Eulerian perturbation of plasma pressure.  $\vec{\xi}$ is the Lagrangian displacement. $\xi_r$ is the radial component of Lagrangian displacement and  $\xi_{\parallel}, \xi_{\perp}$ are the  projections of the Lagrangian displacement in the magnetic surfaces parallel and perpendicular to the magnetic field lines: 



\begin{equation}
\xi_{\perp} = (\xi_{\varphi} B_{z ,0}- \xi_z B_{\varphi,0}) / B, \;\;\xi_{\parallel}  = \vec{\xi}\cdot \vec{B_0} / B _0.
\label{xibotpara}
\end{equation}

\noindent
$ \vec{1}_{\parallel}, \vec{1}_{\perp}$ are the unit vectors in the magnetic surfaces respectively parallel and perpendicular to the magnetic field lines. 

Since the equilibrium quantities are independent of $\varphi$ and $z$ the wave variables can be put proportional to the exponential factor $\exp (i (m\varphi + k_{z}z))$ with  $m,k_z$ the azimuthal and axial wave numbers, $m$ is an integer. For example, for the Lagrangian displacement we write
\begin{equation}
\vec{\xi} (\vec{r};t) =   \vec{\hat{\xi}}(r) \exp (i (m\varphi + k_{z} z - \omega t)).
\label{CylWaves}         
\end{equation}
$\vec{\hat{\xi}}(r)$ is the  radially varying amplitude of $\vec{\xi}$.  In what follows the hat on $\vec{\xi}$ and on the other wave variables will be omitted.  It is convenient to introduce the  wave vector  $\;\vec{k} =(0, m/r,k_{z})$.

The linear MHD waves can be described by two ordinary differential equations for $\xi_r$ and $P'$ \cite[see e.g.][]{appert74,sakurai91,goossens92,goossens95}
\begin{eqnarray}
D\frac{\displaystyle d(r \; \xi_r)}{\displaystyle dr}&  = &  C_1 r \;\xi_r - C_2 r \;P', \nonumber  \\
\mbox{} \nonumber \\
D\frac{\displaystyle d P'}{\displaystyle dr} & = & C_3 \xi_r - C_1 P'. \nonumber 
\label{DiXirP}
\end{eqnarray}
The coefficient functions $D, C_1, C_2, C_3$ are given by \cite{appert74,sakurai91}. The frequencies $\omega_A$ and $\omega_C$ are the local Alfv\'{e}n frequency and the local cusp frequency. They are defined for the planar case in Equations~(\ref{AlfvenFre1}) and (\ref{CuspFre1}). Here in the cylindrical case  their squares are defined as
\begin{equation}
\omega_A^2 = \frac{f_B^2}{\mu \rho_0} = \frac{\displaystyle \left (k_z B_{z,0} + \frac{m}{r} B_{\varphi,0}\right)^2}{ \mu \rho_0}, \;\;
\omega_C^2  = \frac{v_S^2}{v_S^2 + v_A^2} \omega_A^2.
\label{OmegaAC}
\end{equation}
\noindent
$v_A$ and $v_S$ are the Alfv\'{e}n speed and the speed of sound as before and are defined in Equation~(\ref{VAVS}).  In a non-uniform plasma they are functions of position.  The quantities $f_B$ and $g_B$ are defined as


\begin{equation}
f_B = \vec{k} \cdot \vec{B_0} = k_z B_{z,0} + 
\frac{\displaystyle m\; B_{\varphi,0}}{\displaystyle r}, \;\; g_B = (\vec{k} \times \vec{B}_0) \vec{1}_r
= \frac{\displaystyle m \;B_{z,0}}{\displaystyle r} - k_z B_{\varphi,0}.
\label{fBgB}
\end{equation}

Note that $\omega_A$ and $\omega_C$ are function of position. For a given set of wave numbers $(m, k_z)$
$\omega_A$ and $\omega_C$  map out two  ranges of frequencies known as the Alfv\'{e}n continuum and the cusp continuum. 
Recall that for the planar case $\omega_C$ was identified as $\mbox{lim}_{k_x \rightarrow \infty} \omega_{sl}^2 = \omega_C^2$.
With the correspondence from planar coordinates ($x,y,z$) to cylindrical coordinates ($r, \varphi, z$) it follows that the motions associated with $\omega_C$ are rapidly oscillating in the radial direction. 


To emphasize that parallel motions are solely driven by the gradient plasma pressure force the parallel component of the equation of motion is written as 
\begin{equation}
\rho_0 \; \omega^2 \;  \xi_{\parallel} = \frac{\displaystyle i f_B}{\displaystyle B_0} \;\delta p.     
\end{equation}
$\delta p $ is the Lagrangian variation of plasma pressure. 

For the discussion of the mixed properties it is necessary to look at the wave variables  
$\xi_{\perp},\;\xi_{\parallel}, \nabla \cdot \vec{\xi} $  and $(\nabla \times \vec{\xi}) $. 
They are given by expressions in $\xi_r$ and $P'$ and their derivatives. Algebraic expressions for $\xi_{\perp},\;\xi_{\parallel}, \nabla \cdot \vec{\xi} $ 
can be found in e.g. \cite{sakurai91}: 

\begin{eqnarray}
\rho_0 (\omega^2 - \omega_A^2) \xi_{\perp} & = & \frac{i}{ \displaystyle B_0}\; C_A, \nonumber \\
\rho_0 (\omega^2 - \omega_C^2) \xi_{\parallel} & = &
\frac{\displaystyle i f_B}{\displaystyle B_0} \;
\frac{\displaystyle v_S^2}{\displaystyle v_S^2 + v_A^2}\; C_S, \nonumber \\
\nabla \cdot \vec{\xi} & = & \frac{\displaystyle -\omega^2}
{\displaystyle \rho_0\; (v_S^2 + v_A^2)(\omega^2 - \omega_C^2)}  \;C_S.
 \label{WaveVariables}
\end{eqnarray}

\noindent The coupling functions are defined as \citep[see e.g.][]{sakurai91}
\begin{equation}
C_A  =  g_B\; P' - \frac{\displaystyle 2 f_B B_{\varphi,0} B_{z,0}}{\displaystyle \mu r}\; \xi_r, \;\;\;
C_S  =  P'- \frac{\displaystyle 2 B_{\varphi,0}^2 }{\displaystyle \mu r} \;\xi_r.
\label{CAS}
\end{equation}
They are linear combinations of $P'$ and $\xi_r$. Hence in general the coupling is realized through the wave variables 
$P'$ and $\xi_r$. For a straight equilibrium magnetic field with  $B_{\varphi,0} =0$,
the coefficients of $\xi_r$ in $C_A$ and $C_S$  vanish. So for a straight field it is the perturbation of total pressure that causes the mixed properties. In that respect we refer to  the comment by \cite{hasegawa82} (see Section~\ref{sect:straight} below).
 $C_A$ depends on the azimuthal wave number $m$ and the longitudinal wave number $k_z$. $C_S$ on the other hand is independent of the wave numbers $(m,k_z)$. 
The coupling functions  play an essential role for the mixing properties of MHD waves and for resonant absorption. They are called coupling functions for the good reason that they couple the differential equations~(\ref{DiXirP}) for $\xi_r$ and $P'$   to the expressions for all of the remaining wave variables $\xi_{\perp},\;\xi_{\parallel}, \nabla \cdot \vec{\xi} ,\;(\nabla \times \vec{\xi})$. First they couple the differential equations  for $\xi_r$ and $P'$ to the algebraic equations~(\ref{WaveVariables}) for $\xi_{\bot},\;\xi_{\parallel}, \nabla \cdot \vec{\xi}.$  When $C_A \neq 0$ the first equation of~(\ref{WaveVariables}) implies  that  $\xi_{\perp} \neq 0$. Similarly when $C_S \neq 0$ the third equation of~(\ref{WaveVariables}) implies  that 
$\nabla \cdot \vec{\xi} \neq 0$. When in addition to $C_S \neq 0$ also $v_S \neq 0$ it follows that $\xi_{\parallel} \neq 0$.

For completeness we  note that the analysis of mixed properties also applies to incompressible motions.  The relevant equations for  incompressible motions can be found in e.g. \cite{goossens92}. See also \cite{goossens09,goossens12}. Incompressibity means that we take the limit $v_S \rightarrow \infty$  and enforce 
$\nabla \cdot \vec{\xi} = 0$ . The last equation of ~(\ref{WaveVariables}) is identically satisfied. The perturbation of total pressure $P'  \neq 0$ is an unknown function. In what follows we shall mainly deal with compressible motions.

Let us now consider $(\nabla \times \vec{\xi})$.  In  Section~\ref{sec:2}  on  linear MHD waves on a uniform plasma of infinite extent it was  pointed out that a division of linear waves can be based on compression (pressure perturbation), parallel displacement and parallel vorticity. A characterization based on the  components $(\xi_x, \xi_y, \xi_z)$ is in general not possible. When we move from Cartesian geometry to cylindrical geometry the horizontal components $(\xi_x, \xi_y)$ are replaced by the components $(\xi_r,  \xi_{\perp})$ in the planes normal to $\vec{B}_0$ and $\xi_z$ is replaced by the component $\xi_{\parallel}$ parallel to the equilibrium magnetic field.    For a uniform plasma of infinite extent the MHD waves could be divided into incompressible waves that propagate parallel vorticity, i.e. the Alfv\'{e}n waves and  waves that propagate compression/pressure perturbations and have a parallel displacement component i.e. the magneto-sonic waves.  In what follows it will be shown that for a non-uniform plasma MHD waves propagate both compression and parallel vorticity  and have non-zero radial, perpendicular and parallel components of displacement and vorticity.  Expressions for the components of $(\nabla \times \vec{\xi})$ are given in \cite{goossens19}. 



The components of $(\nabla \times \vec{\xi}) $  can be  expressed in terms of $(\xi_r, \xi_{\perp}, \xi_{\parallel})$.  Since $\xi_{\perp}, \xi_{\parallel}$ are expressed in terms of $\xi_r$ and $P'$ it follows that  also the components of $(\nabla \times \vec{\xi}) $ can be  expressed in terms of $\xi_r$ and $P'$. When  $(\xi_r, \xi_{\perp}, \xi_{\parallel})$ are non-zero,  the components of vorticity are in general also non-zero.  All of the wave variables are coupled. The MHD waves have mixed properties, they propagate both pressure perturbations and parallel vorticity and have non-zero radial, perpendicular and parallel  components of displacement and vorticity.  In general all wave variables are non-zero  and can be expressed in terms of $P'$ and $\xi_r$ . A situation in which a subset of the wave variables is not coupled to the other wave variables  is an exception. Such a situation will appear for axisymmetric motions in the presence of a straight field.  The clear division into Alfv\'{e}n waves and magneto-sonic waves  that exists for a uniform plasma of infinite extent does not any  longer hold. 

Hence in general for linear MHD waves on a non-uniform plasma 
\begin{eqnarray}
\xi_r & \neq & 0, \;\;P'\neq 0, \nonumber \\
\xi_{\perp} & \neq & 0, \;\; \xi_{\parallel} \neq 0,\; \nonumber \\
\nabla \cdot \vec{\xi} & \neq & 0, \;\;(\nabla \times \vec{\xi})  \neq 0. 
\label{WaveVariables3}
\end{eqnarray}

\subsection{Axisymmetric motions}

Let us consider the special case of axi-symmetric motions with $m=0$. The expressions for $f_B, \; g_B,\;C_A, \; C_S$ can be simplified to 
\begin{eqnarray}
f_B & =  & k_z B_{z,0},  \;\;\;g_B = -k_z \;B_{\varphi,0},   \nonumber \\
C_A & = & - k_z \;B_{\varphi,0} \{  P'  - 2 \; \frac{B_{z,0}^2\;\xi_r}{\mu\;r}\},
\;\;\; C_S = P' - 2 \frac{B_{\varphi,0}^2}{\mu r} \xi_r.
\label{CASm0}
\end{eqnarray}
For a twisted magnetic field with both a longitudinal component $B_{z,0}$ and an  azimuthal component $B_{\varphi,0}$  the coupling functions $C_A$ and $C_S$  are non-zero.  This implies that  the preceding analysis on mixed properties also applies to axi-symmetric motions. The axi-symmetric motions propagate vorticity and pressure perturbations. The situation is different when the magnetic field is straight.

Since  $C_A$ and $C_S$ are functions of position  the coupling of the equations depends on position and so does the strength of the mixing of the wave properties. For example a  wave  can start off as a predominantly fast wave,   change into a wave that has both fast and Alfv\'{e}n properties and  turn into a predominantly Alfv\'{e}nic wave. MHD waves have mixed properties and have  different appearances in different parts of the plasma because of the  inhomogeneity of the plasma. This phenomenon was discussed by e.g. \cite{goossens02b,goossens08b,goossens11,goossens12,goossens14,goossens19}.  It is remarkable that this behaviour went largely unnoticed when resonant absorption was investigated. It is clearly present in early numerical investigations by  \cite{poedts89,poedts90}. For instance, \cite{poedts89} study the heating of solar coronal loops approximated by straight cylindrical, axisymmetric plasma columns in which incident waves excite oscillations driven by a periodic external driver. The numerical results show that a considerable part of the incident energy is actually dissipated in a localized narrow resonant layer in the radial direction. The real and imaginary parts of the components of the perturbed magnetic field, total pressure perturbation and the divergence of the velocity field, shown in their Figures 3, 4, and 5, display sharp variations in the narrow layer around the resonant surface $r=r_s$.The MHD wave changes from a compressional wave with pressure variations to the right of the resonant position to a wave with almost no pressure variations to the left of the resonant position.  Waves with mixed properties are also referred to as coupled MHD waves. This is a rather strange name as it seems to suggest that there are two or more waves involved. In linear MHD there are infinitely many waves but there is no interaction.

\subsection{Straight equilibrium magnetic field}\label{sect:straight}\label{sec:3.3}

Let us now focus on MHD waves in the presence of a straight field. For a straight field  ($B_{\varphi,0} = 0$) the magnetic surfaces are cylinders: $r = \mbox{constant}$ and the  $\varphi$- and $z$-directions  are the directions in the magnetic surfaces respectively perpendicular and parallel to the magnetic field lines.  The $r$-direction is normal to the magnetic surfaces.   Hence  $\xi_r$ is associated with motions normal or across magnetic surfaces;  $\xi_{\parallel} = \xi_z$ are  motions along the magnetic field lines and $ \xi_{\perp} =\xi_{\varphi}$  are motions in the magnetic surfaces perpendicular to the magnetic field  lines. For a straight field the expressions for  $f_B, \;g_B, \;C_A, \;C_S, $   are simplified to

\begin{eqnarray}
f_B & =  & k_z B_{z,0}, \;\;\;g_B = \frac{m}{r}B_{z,0},  \nonumber \\
C_A & = & g_B \;P' = \frac{m}{r} B_{z,0} \;P', \;\;\; C_S = P'. \
\label{CASBz}
\end{eqnarray}
The  local Alfv\'{e}n and cusp frequency are now independent of the azimuthal wave number $m$

\begin{equation}
 \omega_A^2  = \frac{\displaystyle (\vec{k} \cdot
\vec{B})^2}{\displaystyle \mu \rho} \;=\;k_z^2\;v_A^2 = k_{\parallel} \;v_A^2, \;\;
 \omega_C ^2 = \frac{v_S^2}{v_S^2 +  v_A^2} \omega_A^2. 
\label{AlfvenCuspFreStrF}
\end{equation}
 $v_A$  and $v_S$ are  the Alfv\'{e}n velocity and the velocity of sound, defined in Equation~(\ref{VAVS}). 

The coupling functions $C_A, \;C_S$ only contain $P'$. The coefficients of $\xi_r$ in $C_A$ and $C_S$ vanish when $B_{\varphi,0} = 0$. Hence the coupling of the waves variables is solely due to $P'$ .
This means that the equations are  coupled because of $P'$. This was already known by 
\cite{hasegawa82}. They noted that \begin{quote}The basic characteristic of the ideal Alfv\'{e}n wave is that the total pressure in the fluid remains constant during the passage of the wave as a consequence of the incompressibility condition. For  inhomogeneous medium, however, the total pressure, in general, couples with the dynamics of the motion, and the assumption of neglect of pressure perturbations becomes invalid. \end{quote}
As far as the wave numbers  $(m,k_z)$ are concerned $C_A$ no longer depends on $k_z$, it depends only on $m$.  This can be interpreted that the coupling is due to the azimuthal wave number $m$.

The differential equations (\ref{DiXirP}) for $\xi_r$ and $P'$   and  the algebraic equations (\ref{WaveVariables}) for  $\xi_{\perp},\;\xi_{\parallel}, \nabla \cdot \vec{\xi} $   are now 

\begin{eqnarray}
D\frac{\displaystyle d(r \xi_r)}{\displaystyle dr}&  = &  - C_2 r
P', \nonumber \\
\frac{\displaystyle d P'}{\displaystyle dr} & = &
\rho_0\; (\omega^2 - \omega_A^2) \xi_r, \nonumber \\
\rho_0 \; (\omega^2 - \omega_A^2) \xi_{\varphi} & = & \frac{i m}{r} P',
\nonumber \\
\rho_0 (\omega^2 - \omega_C^2) \xi_{z} & = & i k_{z}
\frac{\displaystyle v_S^2}{\displaystyle v_S^2 + v_A^2} P',
\;\;\;\; \rho_0\; \omega^2 \xi_{z} = i k_{z} \delta p,  \nonumber\\
\nabla \cdot \vec{\xi}& = & \frac{\displaystyle -\omega^2 }
{\displaystyle \rho_0\; (v_S^2 + v_A^2)(\omega^2 - \omega_C^2)} P'.
\nonumber\\
\label{EqStrField1}
\end{eqnarray}


The  equations  for the components of $\nabla \times \vec{\xi}$  are
\begin{eqnarray}
(\nabla \times \xi)_r & = &  k_z  \frac{m}{r} \frac{v_A^2}{v_S^2 + v_A^2} \frac{\omega^2}{\rho_0 (\omega^2 - \omega_A^2)(\omega^2 - \omega_C^2)}P', \nonumber \\
(\nabla \times \xi)_{\varphi} & =  &  - i k_z \frac{\displaystyle d}{\displaystyle dr}\left \{\frac{\displaystyle v_S^2}{\displaystyle v_A^2 + v_S^2}
\frac{\displaystyle 1}{\displaystyle \rho_0 (\omega^2 - \omega_C^2)} \right \}  P' \nonumber \\
 & + & i k_z \frac{\displaystyle \omega^2}{\displaystyle \rho_0 (\omega^2 - \omega_A^2)(\omega^2 - \omega_C^2)}\frac{\displaystyle v_A^2}{\displaystyle v_A^2 + v_S^2} \frac{\displaystyle d P'}{\displaystyle dr},   \nonumber \\
 (\nabla \times \xi)_z  & = &  
  -  i \frac{m}{r} \frac{\displaystyle 1} {\displaystyle \left \{\rho_0 (\omega^2 - \omega_A^2) \right \} ^2}
\frac{\displaystyle d}{\displaystyle dr} \left \{\rho_0 (\omega^2 - \omega_A^2)\right \}  P'.
 \label{EqStrField3}
\end{eqnarray}

The equations that govern  incompressible motions are obtained from~(\ref{EqStrField1}) by taking the limit $v_S \rightarrow \infty$ and imposing
 $\nabla \cdot \vec{\xi} = 0$ 

\begin{eqnarray}
\rho_0 (\omega^2 - \omega_A^2) \frac{\displaystyle d(r \xi_r)}{\displaystyle dr}&  = &  \left \{\frac{m^2}{r^2} + k_z^2 \right \}   r P', \nonumber \\
\rho_0\; (\omega^2 - \omega_A^2) \xi_r & = & \frac{\displaystyle d P'}{\displaystyle dr},  \nonumber \\
\rho_0 \; (\omega^2 - \omega_A^2) \xi_{\varphi} & = & \frac{i m}{r} P',
\nonumber \\
\rho_0 (\omega^2 - \omega_A^2) \xi_{z} & = & i k_{z}  P'.
\nonumber\\
\label{EqStrFieldInComp}
\end{eqnarray}
Equations~(\ref{EqStrFieldInComp}) clearly show that  the coupling of the wave  variables is solely due to $P' \neq 0$ .
Properties of incompressible motions were obtained and discussed by \cite{goossens09} and \cite{goossens12}.

Equations~(\ref {EqStrField1}) and  (\ref {EqStrField3}) govern the MHD waves on a non-uniform straight cylindrical plasma column with a straight magnetic field.  There is a natural subdivision between respectively axi-symmetric motions with $m=0$ and non-axisymmetric motions with $m\neq 0$.  The reason is that the equation for $\xi_{\varphi}$ for $m=0$ is decoupled from the remaining equations. This is also reflected by  the fact that $C_A = 0$   for $m =0 $.

Let us first focus on axi-symmetric motions with $m=0$. 
\begin{equation}
C_A  = 0, \;\; C_S = P'.
\label{CASBz}
\end{equation}
The equation for $ \xi_{\perp} =\xi_{\varphi}$  is  decoupled from the remaining equations
\begin{equation}
\rho_0\; (\omega^2 - \omega_A^2) \; \xi_{\varphi} = 0.
\label{AxiAFSw}
\end{equation}
Equation (\ref {AxiAFSw}) can be satisfied in two ways. First of all by choosing
 \begin{equation}
\omega^2 =  \omega_A^2, \;\; \xi_{\varphi} \neq  0. 
\label{AxiAWStrfield}
\end{equation}
\noindent The second choice is 
 \begin{equation}
\omega^2 \neq   \omega_A^2, \;\; \xi_{\varphi} = 0. 
\label{AxiSFastStrfield}
\end{equation}

The solutions given in~(\ref {AxiAWStrfield}) and (\ref {AxiSFastStrfield})  correspond respectively to the axi-symmetric Alfv\'{e}n waves and the sausage magnetosonic waves.  The axi-symmetric MHD waves are decoupled in sausage magneto-sonic waves and axi-symmetric  Alfv\'{e}n
waves.  The solutions for the axi-symmetric magneto-sonic waves are
\begin{eqnarray}
& & P' \neq  0,\nonumber \\
& & \xi_r  \neq 0 , \;\xi_z \neq 0, \;\xi_{\varphi} =0, \nonumber \\
& & \nonumber \\
& & \nabla \cdot \vec{\xi} = \frac{\displaystyle -\omega^2
P'}{\displaystyle \rho_0 (v_S^2 + v_A^2)(\omega^2 - \omega_C^2)}\neq 0, \nonumber \\
& & \nonumber \\
& & (\nabla \times \vec{\xi})_r = 0, \; (\nabla \times \vec{\xi})_z =0, \nonumber  \\
& & \nonumber \\
& & (\nabla \times \vec{\xi})_{\varphi}  =   - i k_z \frac{\displaystyle d}{\displaystyle dr}\left \{\frac{\displaystyle v_S^2}{\displaystyle v_A^2 + v_S^2}
\frac{\displaystyle 1}{\displaystyle \rho_0 (\omega^2 - \omega_C^2)} \right \}  P' \nonumber \\
 & + & i k_z \frac{\displaystyle \omega^2}{\displaystyle \rho_0 (\omega^2 - \omega_A^2)(\omega^2 - \omega_C^2)}\frac{\displaystyle v_A^2}{\displaystyle v_A^2 + v_S^2} \frac{\displaystyle d P'}{\displaystyle dr}.   \nonumber \\
\nonumber  \\
\label{EqStrField-Sausage-MS}
\end{eqnarray}

The solutions for the axi-symmetric Alfv\'{e}n waves are 
\begin{eqnarray}
& & P' = 0, \nonumber \\
& &  \xi_r = 0, \;\; \xi_z = 0,\;\;\xi_{\varphi} \neq 0, \nonumber \\
& & \nabla \cdot \vec{\xi}  =  0, \nonumber \\
& &  (\nabla \times \vec{\xi})_r = - i k_z \xi_{\varphi}, \;\; (\nabla \times \vec{\xi})_{\varphi} = 0, \;\;
(\nabla \times \vec{\xi})_z = \frac{1}{r}\frac{d}{dr}(r \xi_{\varphi}). \nonumber  \\
\label{EqStrField-AxiS-AW}
\end{eqnarray}

For an axi-symmetric non-uniform 1-dimensional cylindrical plasma this is  the  only case where pure Alfv\'{e}n waves  show up in the analysis. Each magnetic surface oscillates with its own local Alfv\'{e}n frequency. 
In a twisted magnetic field $C_A \neq 0$ for $m=0$ so that the equations are coupled and the corresponding  MHD waves have mixed magneto-sonic and Alfv\'{e}n properties. Also  $C_S \neq 0$ for any azimuthal wave number  $m$. 
The absence of pure Alfv\'{e}n waves in a non-uniform 1-D cylindrical  plasma is in stark contrast to situation of a magnetic cylinder with piece wise constant density and magnetic field. \cite{spruit82} showed that solutions with $\nabla \cdot \vec{v} = 0$ exist for any $m$.  Spruit correctly identified these solutions as Alfv\'{e}n waves. Flow patterns for Alfv\'{e}n waves with $m=0$ and $m=1$  are shown in Figure 1 in \cite{spruit82}. In addition to the Alfv\'{e}n waves there are compressive waves. The fact that pure non-axisymmetric Alfv\'{e}n   waves do not exist in a non-uniform straight plasma cylinder is an illustration of how non-uniformity produces MHD waves with mixed properties. Note that the expressions for the components of vorticity for axi-symmetric magneto-sonic waves can be obtained from (\ref{EqStrField3})  by putting $ m=0. $

Let us now turn  back  to  non-axisymmetric motions with $m \neq 0$. Here all wave variables are coupled and all wave variables are non-zero.  In case of a straight field, it is the non-zero Eulerian perturbation of total pressure $P' \neq 0$ that  produces MHD waves with mixed properties reminiscent of Alfv\'{e}n waves and magneto-sonic waves. 

Special interest goes to the components of $\nabla \times \vec{\xi}$. It is obvious that $ (\nabla \times \vec{\xi})_r \neq 0$  irrespective of whether the equilibrium is non-uniform or not.  The same is true for  $(\nabla \times \vec{\xi})_{\varphi}$. The second term is always non-zero.  The first term is non-zero for a non-uniform equilibrium. For  a piece-wise constant density model the derivative results in a $\delta$-function contribution. The parallel component  
$(\nabla \times \vec{\xi})_z $  is non-zero for a non-uniform equilibrium with 
\begin{equation}
\frac{\displaystyle d}{\displaystyle dr} \left \{\rho_0 (\omega^2 - \omega_A^2)\right \} 
\label{NonUniformOmegaA}
\end{equation}
different from zero. In a fully non-uniform equilibrium this condition is satisfied everywhere. In a piece-wise constant density model the derivative results in a $\delta$-function contribution.

For non-axisymmetric MHD waves on a non-uniform plasma column with a straight magnetic field all wave variables are non-zero and coupled. The coupling factor is $P'$. This means that any  given variable can be expressed  in terms of another wave variable. Let us see what we can do with for example compression  and parallel vorticity. 
Together with the parallel displacement $\xi_z$ these are the two quantities that were used in Section~\ref{aws} to distinguish between Alfv\'{e}n waves and magneto-sonic waves. The expressions for compression  $\nabla \cdot \vec{\xi}$ and for parallel vorticity $(\nabla \times \vec{\xi})_z$  for non-axisymmetric motions in a straight field can be rewritten  in compact form as 
\begin{equation}
\nabla \cdot \vec{\xi} = N_C \;P', \;\;(\nabla \times \vec{\xi})_z = i \;m\; N_V \;P',
\label{CompVorticity1}
\end{equation}
with
\begin{eqnarray}
N_C & = &  \frac{\displaystyle -\omega^2}{\displaystyle \rho_0\; (v_S^2 + v_A^2)(\omega^2 - \omega_C^2)}, 
\nonumber  \\
N_V & =  & \frac{\displaystyle -1}{\displaystyle r  \left \{\rho_0 (\omega^2 - \omega_A^2) \right \} ^2}
\frac{\displaystyle d}{\displaystyle dr} \left \{\rho_0 (\omega^2 - \omega_A^2)\right \}. 
\label{NCNV}
\end{eqnarray}
Equation~(\ref{CompVorticity1}) can then be rewritten as 
\begin{equation}
(\nabla \times \vec{\xi})_z = i \;m\; \frac{N_V}{N_C} \;\nabla \cdot \vec{\xi};
 \;\; \nabla \cdot \vec{\xi} = - \frac{i}{m} \frac{ N_C}{N_V}\;(\nabla \times \vec{\xi})_z. 
\label{CompVorticity2}
\end{equation}
This equation shows that a non-axisymmetric  compressional motion immediately generates vorticity and vice versa a vortical motion generates compression. It is impossible to have one property without the other one. MHD waves that propagate compression but no vorticity or vice versa do not exist. The waves have always mixed properties. 

The cylindrical model with a straight magnetic field has a Cartesian analogue. The Cartesian version has a vertical magnetic field along the $z$-axis and the direction of inhomogeneity along the $x$-axis. The cylindrical case with a straight field and axi-symmetric waves with $m=0$ then corresponds to $k_y =0$.  For $k_y =0$ the Cartesian equations for the wave variables are decoupled in equations for the magneto-sonic waves and equations for Alfv\'{e}n waves. However, for $k_y \neq 0$ the equations are coupled and the MHD waves have mixed properties.  Examples of this behaviour can be found in e.g. \cite{tirry97,tirry97a,degroof00,degroof02b,degroof02a}.

\section{Resonant absorption}\label{sec:4}

\subsection{General discussion}

Resonant absorption is a prominent example of the behaviour of MHD waves with mixed properties. 
The coupling functions   $C_A$ and $C_S$  depend on position. Hence, the strength of the mixing of the wave properties depends on position.  MHD waves have different appearances in different parts of the plasma because of the  inhomogeneity of the plasma. The  phenomenon that the properties of  MHD  waves change as the wave propagates through a non-uniform  environment  is most clearly at work in resonant absorption. For example,  in the case of resonant Alfv\'{e}n waves the MHD wave arrives at  a position  where it  behaves as an almost pure Alfv\'{e}n wave. However, it has pressure variations. Similarly, in the case of resonant cusp waves the MHD wave arrives at a position where it can behave as a slow wave for perpendicular propagation.  Resonant absorption and resonant waves have been discussed previously, see e.g.  \cite{ionson78,hollweg88,goossens11,goossens19,goossens20,goossens21}. 
We now turn to the  discussion of  MHD waves with frequencies in the Alfv\'{e}n continuum and the slow continuum. Resonance absorption can occur in any situation where total pressure fluctuations are imparted to field lines satisfying the Alfv\'{e}n and cusp resonance conditions \citep{hollweg88}. Resonant absorption has a long history in fusion plasma physics, space plasma physics, solar physics and astrophysics.  A characterization was given by  \cite{parker91} who noted that resonant absorption in the Alfv\'{e}n continuum is to be expected when  a wave with a phase velocity $\omega / k$ spans a region in which the variation of the Alfv\'{e}n velocity $v_A$ across the region provides the  resonance condition  $\omega / k = v_A$. Nonuniformity is key  to the process.  The Alfv\'{e}n velocity $v_A$  and  Alfv\'{e}n frequency $\omega_A$ depend on position.  The resonance occurs at the position $r_A$ where the frequency of the wave $\omega$ is equal to the local Alfv\'{e}n frequency  $\omega = \omega_A(r_A)$. This is schematically represented in the top panels of Figure~\ref{fig:f1}. In models with a density jump at the interface between two regions the eigenfrequency of the fundamental kink mode is in between the internal and external Alfv\'en frequencies. When the discontinuous transition is replaced with a continuous variation in a non-uniform transitional layer, the fundamental kink mode has its eigenfrequency in the Alfv\'en continuum and couples to an Alfv\'en continuum mode, leading to energy absorption and damping. The bottom panels of Figure~\ref{fig:f1} represent the situation in cylindrical straight magnetic field models with a non-uniform radial variation of density and Alfv\'en speed. Models that have been widely used to study resonantly damped oscillations in solar coronal loops \citep{goossens11}.

Absorption of energy at the resonant position requires that $P' \neq 0$ there. This does not mean that the wave is an 
acoustic wave. It means that the wave has mixed properties with very strong Alfv\'{e}n characteristics.  There needs to be a connection  to the world outside  the resonant surface for resonant absorption to operate and $P'\neq 0 $ provides this connection.  $P'\neq 0 $ is the simplified condition for a straight field. For a magnetic field with both a longitudinal component $B_{z,0}$ and an azimuthal component $ B_{\varphi,0}$ the condition is $C_A \neq 0$. In the general case $C_A$ provides the link to the outside world.   Axi-symmetric ($m=0$) MHD waves in a straight field do not undergo resonant Alfv\'{e}n wave damping.  However, when $B_{\varphi,0} \neq 0$,  $\xi_r$ does undergo a jump for a wave  with an azimuthal wave number $m=0$.  The jump is proportional to both $B_{\varphi,0}$ and the longitudinal wave number $k_z$. An axi-symmetric wave with its frequency in the Alfv\'{e}n continuum is resonantly damped in an equilibrium with a twisted magnetic field. Examples of absorption of axi-symmetric waves in the Alfv\'{e}n continuum with a twisted field can be found in \cite{poedts89,poedts90}  and \cite{goossensandpoedts92} for the driven problem and in \cite{giagkiozis16} for the eigenvalue problem.

The local Alfv\'{e}n frequency  $\omega_A(r)$ and the local cusp frequency  $\omega_C(r)$  vary with position $r$ and they map out two intervals of frequencies 
\begin{equation}
 \mbox{AC} = [\mbox{min}\;\omega_A(r), \;\;\mbox{max}\;\omega_A(r) \;], \;\;\;
\mbox{SC} =  [\mbox{min}\;\omega_C(r), \;\;\mbox{max}\;\omega_C(r) \;].
\label{ACContinuum}
\end{equation}






In what follows  we consider standing waves. This means that the axial wave number $k_z$ is real and the  frequency $\omega$ is complex  because the waves are damped by resonant absorption. Hence

\begin{eqnarray}
\omega & =&  \omega_R + i \gamma,  \nonumber \\
\exp(- i \omega t)&  = & \exp(- i \omega_R t)\exp(\gamma t) = \exp(- i \omega_R t)\exp(- t/ \tau_d).
\label{Dampedexp}
\end{eqnarray}
For damping in the Alfv\'{e}n continuum 
\begin{equation}
\omega = \omega_R + i \gamma, \;\;  \gamma = \omega_I, \;\;\omega_R = \omega_A(r_A)
\label{ComplexfrequencyAC}
\end{equation}
and for damping in the slow continuum 
\begin{equation}
\omega = \omega_R + i \gamma, \;\;  \gamma = \omega_I, \;\;\omega_R = \omega_C(r_C).
\label{Complexfrequency}
\end{equation}

The real and imaginary parts of the frequency $\omega_R$ and $\gamma$ are related to the period and the damping time $\tau_D$ as
\begin{eqnarray}
\mbox{Period} &=& \frac{2 \pi}{\omega_R}, \;\; \tau_D = \frac{1}{\mid \gamma \mid }, \nonumber  \\ 
\frac{\tau_D}{\mbox{Period}} & = & \frac{1/\mid \gamma \mid} {2 \pi / \omega_R} = \frac{1}{2 \pi} \frac{\omega_R}{\mid \gamma \mid}, \nonumber \\
\frac{\mid \gamma \mid}{\omega_R} & = & \frac{1}{2 \pi} \frac{\mbox{Period}}{\tau_D}.
\label{PeriodDampingTime}
\end{eqnarray}
In case of weak damping
\begin{equation}
\frac{\mid \gamma \mid }{\omega_R} \ll 1, \;\; \frac{\tau_D}{\mbox{Period}} \gg 1.
\label{WeakD}
\end{equation}

The focus is on  the changes in the spatial properties when MHD waves undergo resonant damping in the Alfv\'{e}n/slow continuum.   The analysis is carried out for a  1-D cylindrical  plasma with a straight magnetic field.  The equations that govern the linear motions in that case are given in Section~\ref{sec:3.3} (see Equations~[\ref{EqStrField1}] and  [\ref{EqStrField3}]).
The effect of the damping on the spatial wave variables is determined by using complex frequencies that arise due to the resonant damping.

\subsection{Resonant absorption by Alfv\'{e}n waves}

Let us first consider the Alfv\'{e}n continuum. For a frequency in the Alfv\'{e}n continuum the dispersion relation for Alfv\'{e}n waves is locally satisfied.  Each magnetic surface oscillates at its own Alfv\'{e}n continuum frequency. Let us determine the structure of the MHD wave with a frequency in the Alfv\'{e}n continuum.  The MHD waves live on $[0, +\infty[$. Here we focus on the spatial behaviour close to the singular point  $r = r_A$ where 
$\omega = \omega_A(r_A)$.  We shall find that  the most striking result is the huge spatial variation of the parallel  vorticity. Parallel vorticity vanishes in the uniform part of the equilibrium. However,  once the MHD wave moves into the non-uniform part parallel vorticity explodes to values that are orders of magnitude bigger than those attained by the horizontal  components. In the non-uniform part of the equilibrium plasma the MHD wave is controlled by parallel vorticity and resembles an Alfv\'{e}n wave with the unfamiliar property that it has pressure variations.
In order to keep the mathematical analysis as simple as possible while  still retaining the essential physics we consider a straight magnetic field.

Since 2002 \citep{ruderman02,goossens02a} resonant absorption of kink waves is a popular mechanism for explaining the rapid damping of standing and propagating MHD waves in coronal loops. It also plays a relevant role in turbulent damping models with Kelvin-Helmholtz instability \citep{vandoorsselaere21}. The simple model invokes MHD waves superimposed on a cylindrical plasma column in static  equilibrium. Cylindrical coordinates $(r, \varphi, z)$ are used.  The inhomogeneity necessary for resonant absorption to operate is usually provided by the equilibrium density $\rho_0(r)$ that is constant in the internal and the external part of the loop with values $\rho_i$ and $\rho_e$, but that varies continuously  from  $\rho_i$ to $\rho_e$ in a transitional layer $[R - l/2, \;\;R + l/2]$  of thickness $l$ (see Figure~\ref{fig:f1}).  The fundamental radial mode of  kink ($m=1$)  waves has its frequency in the Alfv\'{e}n continuum and is always  resonantly damped. Here we consider standing waves.  This  means that the axial wave number $k_z$ is real and the  frequency $\omega$ is complex. 


Analytical expressions for the period and damping time have been obtained by  the use of the  thin tube and  thin boundary TTTB  approximation so that  $k_z R  \ll 1, \; l/R  \ll 1$. Analytical expressions for the damping time $\tau_D$   were derived by \cite{hollweg90a,goossens92,ruderman02,vandoorsselaere04,arregui19}. Often simple analytical prescriptions for the density profile were used.





Recall from Section~\ref{sec:3} that all wave variables are non-zero in a non-uniform plasma.   There are  no pure magneto-sonic waves and no pure Alfv\'{e}n waves  in a non-uniform plasma.   For a straight constant field $\vec{B}_0 = B_0 \vec{1}_z $ the spatial variation of $\omega_A^2$ is solely due to the spatial variation of the equilibrium density $\rho_0$. Hence

\begin{equation}
\frac{\displaystyle d\rho_0}{\displaystyle dr} \neq 0
\label{NonDensity}
\end{equation}
is the important quantity for the resonant behaviour.  For a constant straight magnetic field the plasma pressure $p_0$ is also constant and $\nabla p_0 = 0$. Hence

\begin{eqnarray}
\rho_0\; v_A^2 & = & \frac{B_0^2}{\mu_0} = \mbox{constant},\;
\rho_0 \; \omega_A^2 = k_z ^2 \frac{B_0^2}{\mu_0} = \mbox{constant}, \nonumber \\
\frac{\displaystyle v_S^2}{\displaystyle v_A^2 + v_S^2} & = & 
\mbox{constant}, \;  \rho_0 \;\omega_C^2 =  \rho_0 \; \frac{v_S^2}{v_S^2 + v_A^2} \;  \omega_A^2\;= \; \mbox{constant}.
\label{ConstantAC}
\end{eqnarray}
The expression for the parallel component of vorticity then simplifies to 

\begin{eqnarray}
(\nabla \times \vec{\xi})_{\parallel} = (\nabla \times \xi)_z & = & -  
i \frac{m}{r} \frac{\displaystyle  \omega^2 } {\displaystyle \left \{\rho_0 (\omega^2 - \omega_A^2) \right \} ^2}\;
\frac{\displaystyle d  \rho_0}{\displaystyle dr} \; P'.
\label{vorticityZ1}
\end{eqnarray}

\subsection{Analysis close to the ideal Alfv\'{e}n resonant point}

This section investigates the behaviour of the wave variables for a standing damped resonant wave close to the ideal resonant point $r_A$.  For a resonantly damped standing wave we use  Equation~({\ref{ComplexfrequencyAC}) and the weak damping approximation~({\ref{WeakD}). \cite{soler13} computed in  non-stationary MHD  the components of the Lagrangian displacement $\xi_r, \;\;\xi_{\varphi}$ and the Lagrangian perturbation of total pressure $P'$.  We shall present the components of vorticity  in Figures~\ref{fig:f3}--\ref{fig:f5}. The parallel vorticity was computed in non-stationary resistive MHD and is shown in their Figure 4 by \cite{goossens12}. The aim here is to obtain a simple understanding of how the fundamental quantities compression and vorticity are influenced by the non-stationary behaviour of the resonantly damped wave. Hence the analytic investigation that follows.

Recall that the  fundamental conservation law for resonant Alfv\'{e}n waves for a straight field was obtained by \cite{sakurai91}
\begin{equation}
 P' = \mbox{constant}, \;\; [P'] = 0.
\label{CAstraightconstant}
\end{equation}

In view of the observed periods and damping  times, the weak damping limit~(\ref{WeakD}) is an accurate approximation. In what follows 
 we use  approximate  results by retaining the dominant terms, i.e. the lowest order terms in $\mid \gamma \mid / \omega_R $:

\begin{eqnarray}
 \omega^2 \frac{d \rho_0}{dr}  & = & \omega_R^2 (1 + i \frac{\gamma}{\omega_R})^2\; \frac{d \rho_0}{dr}  \; \approx \;  \omega_R^2 \frac{d \rho_0}{dr}, \nonumber  \\
&& \nonumber \\ 
\omega^2 - \omega_A^2 & \approx & 2 i\; \gamma \;\omega_R, \nonumber  \\
&& \nonumber \\
\omega^2 - \omega_C^2 & \approx & \omega_A^2  \frac{v_A^2 }{v_S^2  + v_A^2}, \nonumber \\
&& \nonumber \\
\frac{1}{\rho_0 (\omega^2 - \omega_A^2 )} & \approx &  i \frac{\pi}{\omega_R^2 \; \rho_0(r_A)}
 \left\{ \frac{ \tau_D }{ \mbox{Period}}\right\},
\nonumber \\
&& \nonumber \\ 
-  i \frac{m}{r} \frac{\displaystyle \omega^2} {\displaystyle \left \{\rho_0 (\omega^2 - \omega_A^2) \right \} ^2}
\frac{\displaystyle d \rho_0}{\displaystyle dr}  & \approx  & 
 i \frac{m}{r_A}\; \frac{1}{ \rho_0(r_A)\; \omega_R^2} \; \frac{1}{\rho_0(r_A)}\frac{d \rho_0}{dr} \; \pi ^2\; 
  \left\{ \frac{ \tau_D }{ \mbox{Period}}\right\}^2.  \nonumber \\
& &
\label{IResults}
\end{eqnarray}
With the use of (\ref{IResults}) the components of the displacement at $r=r_A$ are

\begin{eqnarray}
\xi_{z} & =  &  i k_z \; \frac{v_S^2}{ v_A^2}  \; \frac{1}{\rho_0 \omega_R^2}  P',   \nonumber \\
\xi_{\perp} & = & \xi_{\varphi} \approx \;  - \frac{ i m/r_A}{\rho_0(r_A)} \;\frac{\pi}{\omega_R^2}
\; \left\{ \frac{ \tau_D }{ \mbox{Period}}\right\}  P',  \nonumber \\
& &  \nonumber \\
 \xi_r & \approx &i \frac{\pi}{\omega_R^2 \; \rho_0(r_A)} \; \left\{ \frac{ \tau_D }{ \mbox{Period}}\right\}
 \frac{d P'}{dr}, \nonumber \\
\label{CompDispl}
\end{eqnarray}

\noindent and for the components of vorticity $(\nabla \times \vec{\xi})$ we have

\begin{eqnarray}
(\nabla \times \vec{\xi})_r   & \approx & -i \;  \frac{m}{k_z \;r_A}   \pi 
\frac{1}{\; B_0^2 / \mu_0}
\left\{\;\frac{\tau_D}{\mbox{Period}} \right\}\;   \; P', 
\nonumber \\
& &  \nonumber \\
(\nabla \times \vec{\xi})_{\varphi}   & \approx &  \frac{1}{\pi \;  k_z}\;
\frac{d P'}{dr} \frac{1} {B_0^2 / \mu_0}  \left\{\frac{\tau_D}{\mbox{Period}}\right\},  \nonumber  \\
& &  \nonumber \\
 (\nabla \times \vec{\xi})_z & \approx  &  i \frac{m}{r_A}\; \frac{ \pi^2}{k_z ^2} \; \frac{1}{\rho_0(r_A)}\frac{d \rho_0}{dr} 
\;\frac{1}{B_0^2 /\mu_0} \; \left\{ \frac{  \tau_D }{ \mbox{Period}}\right\}^2  \;P'. \nonumber \\
& &  \nonumber \\
\label{ParallelVCompression}
\end{eqnarray}

Now recall that 
$$
\nabla \cdot \vec{\xi} =  \frac{\displaystyle -\omega^2 }
{\displaystyle \rho_0\; (v_S^2 + v_A^2)(\omega^2 - \omega_C^2)} P'
$$
and realize that Equation ~(\ref{ParallelVCompression}) are  relations  between the components of vorticity  and  compression.  In particular the relation between  parallel vorticity and compression is remarkable.  For a uniform plasma of infinite extent these two quanties are mutually exclusive and characterize Alfv\'{e}n waves and fast magneto-sonic waves respectively. Equation~(\ref{ParallelVCompression}) shows that  in a non-uniform plasma there are  no pure magneto-sonic waves and no pure Alfv\'{e}n waves  as they appear in a uniform plasma of infinite extent.

\cite{goossens20} used  a prescription   for the variation of the equilibrium density in the non-uniform layer of thickness $l$ and a simple formula  for  $\tau_D/P$  to make the dependence of   $ \mid (\nabla \times \vec{\xi})_z \mid $ on $k_z R$, $l/R$ explicit. See that paper  for the details.

What about $ d P'/ dr$? For a straight field  the Eulerian perturbation of total pressure is a conserved quantity \citep[see e.g.][]{sakurai91,tirry96,soler13}  and does not undergo strong spatial variations.  This is also verified by numerical computations. This  leads us  to use as an  estimate for  the derivative 
$$
\frac{d P'}{dr} \approx P'/R .
$$
 In Figure~\ref{fig:f2} $P'$ is plotted as  function of $r$  for the fundamental radial mode. This figure shows that a linear function is a very good representation for $P'$  in the region $r \in [0, R]$.  With this approximation we obtain
$$
(\nabla \times \vec{\xi})_{\varphi}   \approx  \frac{1}{\pi \;  k_z  R }\;
 P' \; \frac{1} {B_0^2 / \mu_0}  \left\{\frac{\tau_D}{\mbox{Period}}\right\}.  \nonumber  \\
$$

Recall that these expressions only apply in the non-uniform part of the loop and strictly speaking only close to the position $r_A$ where $\omega = \omega_A(r_A)$. Anyhow these expressions give us a good description  for understanding what happens with the spatial solutions when a wave undergoes resonant damping.

\subsection{Global spatial variation of an MHD wave that undergoes resonant Alfv\'{e}n wave damping}\label{global}

The   fundamental conservation law for resonant Alfv\'{e}n waves is now 
\begin{equation}
 P' = \mbox{constant}, \;\; [P'] = 0.
\label{CAstraightconstant}
\end{equation}

Here we apply our results from the previous two subsections to understand  the spatial  behaviour of MHD waves that undergo resonant absorption in the equilibrium coronal loop defined in Section 3.  Let us recall  that  we have repeated the computations by \cite{soler13} in  non-stationary MHD. In the original publication the attention went to the components of Lagrangian displacement $\xi_r, \xi_{\varphi}$ and the Lagrangian perturbation of total pressure $P'$.  Based on the simple relations presented in (\ref{CompDispl}) we arrive at the following ordering

\begin{equation}
\mid  \xi_z \mid   \ll  \mid  \xi_r \mid  \ll    \xi_{\varphi} \mid , \;\;\; \vec{\xi} \approx \xi_{\varphi} \vec{1}_{\varphi}. 
\label{RAWDisplacement}
\end{equation}

Similarly, for the components of vorticity  Equation~(\ref{ParallelVCompression}) tells us that

\begin{equation}
 \mid( \nabla \times \vec{\xi})_{\varphi} \mid  \approx \mid( \nabla \times \vec{\xi})_r \mid  \ll 
\mid( \nabla \times \vec{\xi})_z \mid , \;\;\;  \nabla \times \vec{\xi} \approx ( \nabla \times \vec{\xi})_z \vec{1}_z.
\label{RAWDVorticity}
\end{equation}

It is instructive to compare Equations~(\ref{RAWDisplacement}) and~(\ref{RAWDVorticity}) to the results obtained in Section~\ref{sec:2} for Alfv\'{e}n waves in a uniform plasma of infinite extent  in the limit  $k_x \rightarrow \infty$.  The obvious conclusion is that at  and in the close vicinity of the resonant point the wave is a  resonant Alfv\'{e}n wave with the strange property that it has pressure perturbations. The analytic considerations are confirmed by the numerical calculations.  The components of vorticity have now been computed and are presented in Figures~\ref{fig:f2}--\ref{fig:f4}.  Parallel vorticity  in non-stationary resistive MHD was computed  by \cite{goossens12}. 

Compression and horizontal vorticity do not require non-uniformity. They are non-zero everywhere in the plasma column. However, parallel vorticity is zero in a uniform plasma. It requires a spatial variation of the equilibrium magnetic field and/or the equilibrium density. For a constant axial magnetic field it is the spatial variation of density that causes the non-zero parallel vorticity. Hence for the resonant damping of transverse waves in the Alfv\'{e}n continuum in the equilibrium model of Section 3 we arrive at the following situation.  In the uniform internal part of the loop  $ 0 \leq r \leq R - l/2 $ and in the uniform external part of the loop $ R + l/2  \leq  r < \infty$  there is compression and  horizontal vorticity. There is no parallel vorticity. In the non-uniform part of the loop there is compression and both horizontal and parallel vorticity. At and in the vicinity of the resonant point 
 $ r = r_A $ the parallel vorticity dominates  over the horizontal vorticity  and the MHD wave is almost a pure Alfv\'{e}n wave. 
 
In Figures~\ref{fig:f2} and \ref{fig:f3}  we have plotted compression and the components of vorticity computed on the interval $[0, \;+\infty[$ for $l/R= 0.5,\;\; k_z R = 0.1, \;\;\zeta = \rho_i/\rho_e= 3.$  The normalization 
$$
\mbox{max}\left\{\mid \nabla \cdot \vec{\xi} \mid \right \} = 1
$$
has been used in Figures~\ref{fig:f2} and \ref{fig:f3}  and in the following figures. 
Figure~\ref{fig:f2} shows that nothing particular happens for compression. The behaviour does not substantially differ from that found in the stepwise constant case. Figure~\ref{fig:f3} shows that the horizontal components of vorticity are non-zero in the whole domain; the parallel component of vorticity is only non-zero in the non-uniform part of the loop. The parallel component of vorticity is far larger than the horizontal components. In Figure~\ref{fig:f3} the maximal value  of the parallel component is at least 4 orders of magnitude bigger than the maximal values of the horizontal components.  In Figures~\ref{fig:f4} and \ref{fig:f5} we have plotted the components of vorticity for $\zeta = 2$ and three values of $k_z R = 0.1, 0.3, 0.5$  (Figure~\ref{fig:f4}) and for $k_z R = 0.1$ and three values of $\zeta =2, 5, 10$ (Figure~\ref{fig:f5}). In all cases $l/R = 0.5$. Since the parallel vorticity is only non-zero in the non-uniform transitional layer the components have been plotted on the interval $[ R - l/2, \;\;R + l/2]$. Resonant absorption is a clear example of  the phenomenon  that the properties of an MHD wave change when it travels through an inhomogeneous plasma.   In case of resonant Alfv\'{e}n waves the MHD wave arrives at  a position  where it  behaves as an almost pure Alfv\'{e}n wave.

In linear MHD there is no interaction between waves and the behaviour that we discussed is associated with a single MHD wave that lives in the whole space. Pure Alfv\'{e}n waves and pure magneto-sonic waves exist in a uniform plasma of infinite extent. In a non-uniform plasma the MHD waves  combine the properties of the classic Alfv\'{e}n waves and of the magneto-sonic waves in a uniform plasma of infinite extent.   In a non-uniform plasma  MHD waves  propagate both compression and parallel and horizontal vorticity. The properties of the MHD wave depend on the properties of the background plasma.  Hence as an MHD waves propagates through a non-uniform plasma its properties change.   When an MHD  wave moves from a uniform into a non-uniform plasma it is transformed from a fast magneto-sonic wave into a  mixed fast - Alfv\'{e}n  wave  and eventually, when it reaches the resonant point,  into an  Alfv\'{e}n wave. However,  the total pressure perturbation and compression are non-zero everywhere.

\subsection{Resonant absorption by slow waves}

Let us now turn to the slow continuum. The analysis for a frequency in the slow continuum parallels  that for Alfv\'{e}n waves \citep[see e.g.][]{sakurai91,goossensandruderman95}.  The MHD waves live on $[0, + \infty[$ . Here we focus on the spatial behaviour close to the singular point $r = r_C$ where $\omega = \omega_C(r_C)$.   

There are two differences with the Alfv\'{e}n continuum.  Firstly,   plasma pressure is essential  with $v_S^2 \neq 0$ since there are no slow waves for a pressureless plasma. Secondly, contrary to Alfv\'{e}n waves,  for a straight field  there is resonant absorption in the slow continuum for axi-symmetric waves with $m=0$ in the same way as for non-axisymmetric waves. The aspects on slow resonant waves presented in this subsection were discussed by \cite{goossens21}.

Again we  focus on  the popular equilibrium model with a straight magnetic field.  The differential equations for $\xi_r$ and $P'$   and  the algebraic equations for  $\xi_{\perp},\;\xi_{\parallel}, \nabla \cdot \vec{\xi} $  can be found in (\ref{DiXirP}) and (\ref{EqStrField1}). 
The expressions for the components of $\nabla \times \vec{\xi}$ are given in  (\ref{EqStrField3}). 
Absorption of energy at the resonant position requires that $P' \neq 0$ there.  There needs to be a connection  to the world outside  the resonant surface for resonant absorption to operate and $P'\neq 0 $ provides this connection.  $P'\neq 0 $ is the simplified condition for a straight field.
For a magnetic field with both a longitudinal component $B_{z,0}$ and an azimuthal component $ B_{\varphi,0}$ the condition is  $C_S \neq 0$. The jumps $[\xi_r]$ and $\mbox{[}P'\mbox{]}$ are proportional to  $C_S$.  In the general case $C_S$ provides the link to the outside world.


From the discussion on resonant Alfv\'en waves we know that

\begin{equation}
\frac{\displaystyle d}{\displaystyle dr} \left \{\rho_0 (\omega^2 - \omega_A^2)\right \} \neq 0
\label{NonUniformOmegaA}
\end{equation}
 is an  important quantity  for  waves that are resonantly damped in the Alfv\'{e}n continuum, especially the parallel component of vorticity. 
For damping in the slow continuum the  parallel component of vorticity does not play any role of significance. We  shall point out that   the perpendicular component is the dominant component of the vorticity. In particular  the first term in the right side of  $(\nabla \times \vec{\xi})_{\varphi} $ in Equation~(\ref{EqStrField3}) will be important. This term is non-zero when

\begin{equation}
\frac{\displaystyle d}{\displaystyle dr} \left \{\rho_0 (\omega^2 - \omega_C^2)\right \} \neq 0.
\label{NonUniformOmegaC}
\end{equation}

Since, by assumption, $B_z = \mbox{constant}$ the spatial variation of $\omega_A^2$  and  $\omega_C^2$ 
is solely due to the spatial variation of the equilibrium density $\rho_0$.   Hence 

\begin{equation}
\frac{\displaystyle d\rho_0}{\displaystyle dr} \neq 0
\label{NonDensity}
\end{equation}
is the important quantity for the resonant behaviour in the Alfv\'{e}n continuum and in the slow continuum. 

The parallel component of vorticity for a constant axial magnetic field is given in Equation~(\ref{vorticityZ1}).
The perpendicular component in that case is  
\begin{eqnarray}
 (\nabla \times \vec{\xi})_{\perp} =(\nabla \times \vec{\xi})_{\varphi}  & = &  
 i k_z  \frac{v_S^2}{v_S^2 + v_A^2} \frac{\displaystyle  \omega^2 } {\displaystyle \left \{\rho_0 (\omega^2 - \omega_C^2) \right \} ^2}\; \frac{\displaystyle d  \rho_0}{\displaystyle dr} \; P'
  \nonumber  \\
& + & i k_z \frac{\displaystyle \omega^2}{\displaystyle  (\omega^2 - \omega_A^2)} \;
\frac{\displaystyle v_A^2}{\displaystyle v_A^2 + v_S^2}  \;
\frac{\displaystyle 1}{\displaystyle \rho_0 (\omega^2 - \omega_C^2)}
\frac{\displaystyle d P'}{\displaystyle dr}.
\label{vorticityPerp1}
\end{eqnarray}
The first term in the right hand side of  $(\nabla \times \vec{\xi})_{\varphi} $  is the dominant term for frequencies in the slow continuum,  as we shall explain in the next section.  It is  remarkably similar to~(\ref{vorticityZ1}). 

Recall that  $P'$ is  related to  $\nabla \cdot \vec{\xi}$ by an algebraic equation. Hence  the components of vorticity can be in terms of $\nabla \cdot \vec{\xi}$.  Again that parallel/perpendicular vorticity and compression go together in a non-uniform plasma.
 
\subsection{Analysis of slow resonant waves close to the resonant point}

We consider a damped wave with a complex frequency 
that undergoes resonant damping in the cusp continuum and use ({\ref{Complexfrequency}) and adopt the weak damping approximation (\ref{WeakD}).
 We focus  on the position $r_C$ where the ideal resonance occurs $\omega^2 = \omega_C^2 (r_C)$. In what 
follows we simplify the expressions for the wave variables by retaining only the dominant terms in 
$\mid \gamma \mid /\omega_R$. The components of displacement $ \xi_r, \; \xi_{\perp} = \xi_{\varphi}$ and 
the parallel vorticity  $(\nabla \times \vec{\xi})_z$  are  independent of $( \tau_D /  \mbox{Period}).$   These quantities do not feel the cusp resonance. The results for the components of $\vec{\xi}$ and for compression are 





\begin{eqnarray}
\xi_r & \approx &  -\frac{1}{\rho_0}\; \frac{v_S^2}{ v_A^2}\; \frac{1}{ \omega_R^2}  \frac{d P'}{dr}, \nonumber  \\
&& \nonumber \\ 
\xi_{\perp} & \approx & \xi_{\varphi} = - i \frac{ m}{r}\;\frac{1}{\rho_0}\; \frac{v_S^2}{ v_A^2}\; \frac{1}{ \omega_R^2}  P',
\nonumber \\
\xi_z &\approx &  k_z\; \frac{v_S^2}{v_S^2 + v_A^2}\; \frac{\pi} {\rho_0  \; \omega_R^2}  \;
\; \left \{\frac{ \tau_D }{ \mbox{Period}}\right \}\; P',  \;\nonumber \\
&& \nonumber \\ 
\nabla \cdot \vec{\xi} & \approx  & - i \frac{\pi} {\rho_0  \;(v_S^2  + v_A^2) }  \;
\left \{\frac{  \tau_D }{ \mbox{Period}} \right \} P'. \nonumber \\
\label{DisplCompression}
\end{eqnarray}

The components of vorticity are
\begin{eqnarray}
(\nabla \times \vec{\xi})_r & \approx  & - i \; \pi \; k_z \; \frac{m}{r}\; \frac{v_S^2}{v_S^2 + v_A^2}\; 
\frac{1 }{\rho_0 \; \omega_R^2} \;
\left \{\frac{ \tau_D }{ \mbox{Period}}\right \} \;P',  
\nonumber \\
(\nabla \times \vec{\xi})_{\varphi}  & \approx &   
-  i k_z \; \pi ^2 \frac{\displaystyle v_S^2}{\displaystyle v_A^2 + v_S^2}
\frac{1 }{\rho_0 \; \omega_R^2} \; \frac{1}{\rho_0} \;\frac{d \rho_0}{dr}\;
\left\{ \frac{  \tau_D }{ \mbox{Period}}\right\}^2 \; P', \nonumber \\
& + & k_z \pi \;\frac{1}{\rho_0 \; \omega_R^2} \;
 \frac{\displaystyle v_S^2}{\displaystyle v_A^2 + v_S^2}\; 
\left \{  \frac{ \tau_D }{ \mbox{Period}} \right \}\;  \frac{\displaystyle d P'}{\displaystyle dr}, 
\nonumber \\
 (\nabla \times \vec{\xi})_z  & \approx  & - i \frac{m}{r} \;  \left(\frac{v_S ^2}{v_A^2} \right)^2 \; \frac{1}{\rho_0}\; 
\frac{\displaystyle d \rho_0}{\displaystyle dr} \frac{1 }{\rho_0 \; \omega_R^2} \;P'.   \nonumber \\
\label{VorticityRSlow}
\end{eqnarray}

Let us now look at the case of axi-symmetric (sausage) slow waves. There is nothing special about axi-symmetric slow waves.  All we have to do is put $m=0$ in the previous equations and expressions. We note  that $\xi_{\varphi},\; (\nabla \times \vec{\xi})_r, \;(\nabla \times \vec{\xi})_{z}  $ are proportional to $m$ and vanish for $m=0$. Similarly the expressions for $\xi_r, \xi_z, \;\nabla \cdot \vec{\xi}, \;  (\nabla \times \vec{\xi})_{\varphi} $ do not depend explicitly on $m$. They depend implicitly on $m$ since $P'$ depends on $m$. Anyhow the expressions  at the ideal resonant position   for the components of the Lagrangian displacement and $\nabla \cdot \vec{\xi}$  are 

\begin{eqnarray}
\xi_r & = &  -\frac{1}{\rho_0}\; \frac{v_S^2}{ v_A^2}\; \frac{1}{ \omega_R^2}  \frac{d P'}{dr}, \nonumber  \\
&& \nonumber \\ 
\xi_{\perp} & = &\xi_{\varphi} = 0, \nonumber \\
\xi_z &\approx &  k_z\; \frac{v_S^2}{v_S^2 + v_A^2}\; \frac{\pi} {\rho_0  \; \omega_R^2}  \;
\left \{\frac{ \tau_D }{ \mbox{Period}} \right \}\; P' , \;  \nonumber \\
\nabla \cdot \vec{\xi} & \approx  & - i \frac{\pi} {\rho_0  \;(v_S^2  + v_A^2) }  
\left \{\frac{  \tau_D }{ \mbox{Period}}\right \}  P' .
\label{AxiDisCom}
\end{eqnarray}

The components of the vorticity  $\nabla \times \vec{\xi}$  are

\begin{eqnarray}
(\nabla \times \vec{\xi})_r & = & 0, \nonumber \\
(\nabla \times \vec{\xi})_{\varphi}  & \approx &   
-  i k_z \; \pi ^2 \frac{\displaystyle v_S^2}{\displaystyle v_A^2 + v_S^2}
\frac{1 }{\rho_0 \; \omega_R^2} \; \frac{1}{\rho_0} \;\frac{d \rho_0}{dr}\;
\left\{ \frac{  \tau_D }{ \mbox{Period}}\right\}^2 \; P' \nonumber \\
& + & k_z \pi \; \frac{ \tau_D }{ \mbox{Period}}\;\frac{1}{\rho_0 \; \omega_R^2} \;
 \frac{\displaystyle v_S^2}{\displaystyle v_A^2 + v_S^2}\;  \frac{\displaystyle d P'}{\displaystyle dr},
\nonumber \\
(\nabla \times \vec{\xi})_{z}   & = & 0. 
\label{Vorticity}
\end{eqnarray}

Based on the simple relations presented in (\ref{DisplCompression}) and  (\ref{AxiDisCom})
 we arrive at the following ordering

\begin{equation}
\mid  \xi_{\varphi} \mid   \ll  \mid  \xi_r \mid  \ll   \mid  \xi_z \mid , \;\;\; \vec{\xi} \approx \xi_z \vec{1}_z.
\label{RCDisplacement}
\end{equation}

Similarly for the components of vorticity we use  Equations (\ref{VorticityRSlow}) and   (\ref{Vorticity}) to arrive at

\begin{equation}
 \mid( \nabla \times \vec{\xi})_{z} \mid  \approx \mid( \nabla \times \vec{\xi})_r \mid  \ll 
\mid( \nabla \times \vec{\xi})_{\varphi} \mid , \;\;\;  \nabla \times \vec{\xi} \approx ( \nabla \times \vec{\xi})_{\varphi}\vec{1}_{\varphi}.
\label{RCDVorticity}
\end{equation}

It is instructive to compare Equations~(\ref{RCDisplacement}) and~(\ref{RCDVorticity}) to the results obtained in Section~\ref{sec:2} for slow  waves in a uniform plasma of infinite extent  in the limit  $k_x \rightarrow \infty$ .  The obvious conclusion is that at  and in the close vicinity of the resonant point the wave is the non-uniform version of the slow wave 
we discussed there  in the  $k_x \rightarrow \infty$ . The analytic considerations are confirmed by the numerical calculations.  

The conclusion of our discussion of slow resonant waves is that $\xi_z$, $\nabla \cdot \vec{\xi}$ and $(\nabla \times \xi)_{\varphi} $ are the dominant quantities:

\begin{equation}
\vec{\xi}  \approx  \xi_z \vec{1}_z, 
\;\; \nabla \times \vec{\xi}   \approx  (\nabla \times \vec{\xi})_{\varphi}\;  \vec{1}_{\varphi}.
\label{DominantB}
\end{equation}
The most striking result are the huge spatial variations of the parallel displacement, compression and  perpendicular vorticity. These three quantities are non-zero in the entire equilibrium configuration, but in  the non-uniform part  their values are far  bigger than those attained in the uniform plasma. In the non-uniform part of the equilibrium plasma an MHD wave with a frequency in the cusp continuum is very compressible as expected but also  carries a huge amount of perpendicular  vorticity. In addition the results for these three quantities are in first order independent of the wave number $m$.

\section{Conclusions}\label{sec:conclusions}
In a uniform plasma of infinite extent MHD  waves can be subdivided in two classes. The first class contains incompressible Alfv\'en waves which are driven by magnetic tension and propagate parallel vorticity. The second class contains the compressive (fast and slow) magneto-sonic waves, which are driven by both pressure perturbations and magnetic tension. The waves have either parallel vorticity and no pressure perturbations and no parallel displacement (Alfv\'en waves),  or they have pressure variations and parallel displacement and no parallel vorticity (magneto-sonic waves). This clear division is no longer present in non-uniform plasmas and the MHD waves have now mixed properties. In a given
part of the equilibrium an MHD wave can strongly resemble a magneto-sonic wave with little or no resemblance to 
Alfv\'en waves; while in another part of the equilibrium the MHD wave is practically an Alfv\'en wave,  which has the amazing property of being accompanied by variations in pressure. We can therefore have waves with pressure variations and parallel vorticity at the same time. Examples have been presented in which this phenomenon is prominent in the context of resonant absorption by Alfv\'en waves and by slow waves. 

\section*{Acknowledgments}
MG acknowledges the travel grant awarded by the FWO-Vlaanderen that made his participation in the 
2024-AAPPS-DPP2024 conference in Malacca possible. MG presented there an oral contribution ``MHD waves with mixed properties / Alfv\'{e}n waves with pressure variations"  in the session chaired by Prof. Dr. Peng Fei Chen. We are grateful to Prof. Dr. P.F.  Chen for his suggestion to publish a review paper on the subject of that  presentation in RMPPP. We are grateful to Prof. Dr. Mitsuru Kikuchi  for inviting us to submit a special topics  paper with the title suggested by Prof. Dr. P.D.  Chen ``MHD waves with mixed properties / Alfv\'{e}n waves with pressure variations: a review".  This review is an extended version of the invited talk.
The content of the present paper closely follows the actual presentation. It is largely based on published work by the authors and their collaborators. Sections~\ref{sec:2}-\ref{sec:3} borrow from \cite{soler24} and Section~\ref{sec:4} borrows from \cite{goossens20,goossens21}. Here the material is presented in a structured and logical manner.
IA acknowledges support from project PID2024-156538NB-I00 funded by MCIN/AEI/10.13039/501100011033 and by “ERDF A way of making Europe”. RS and JT acknowledge support by the I+D+i project PID2023-147708NB-I00 funded by MICIU/AEI/10.13039/501100011033/  and by  FEDER, EU. TVD was supported by the C1 grant TRACEspace of Internal Funds KU Leuven, and a Senior Research Project (G088021N) of the FWO Vlaanderen. Furthermore, TVD received financial support from the Flemish Government under the long-term structural Methusalem funding program, project SOUL: Stellar evolution in full glory, grant METH/24/012 at KU Leuven. The research that led to these results was subsidised by the Belgian Federal Science Policy Office through the contract B2/223/P1/CLOSE-UP. It is also part of the DynaSun project and has thus received funding under the Horizon Europe programme of the European Union under grant agreement (no. 101131534). Views and opinions expressed are however those of the author(s) only and do not necessarily reflect those of the European Union and therefore the European Union cannot be held responsible for them.

\section*{Conflict of interest}
On behalf of all authors, the corresponding author states that there is no conflict of interest.


\newpage

\begin{figure}
\includegraphics[scale=0.45,angle=-90]{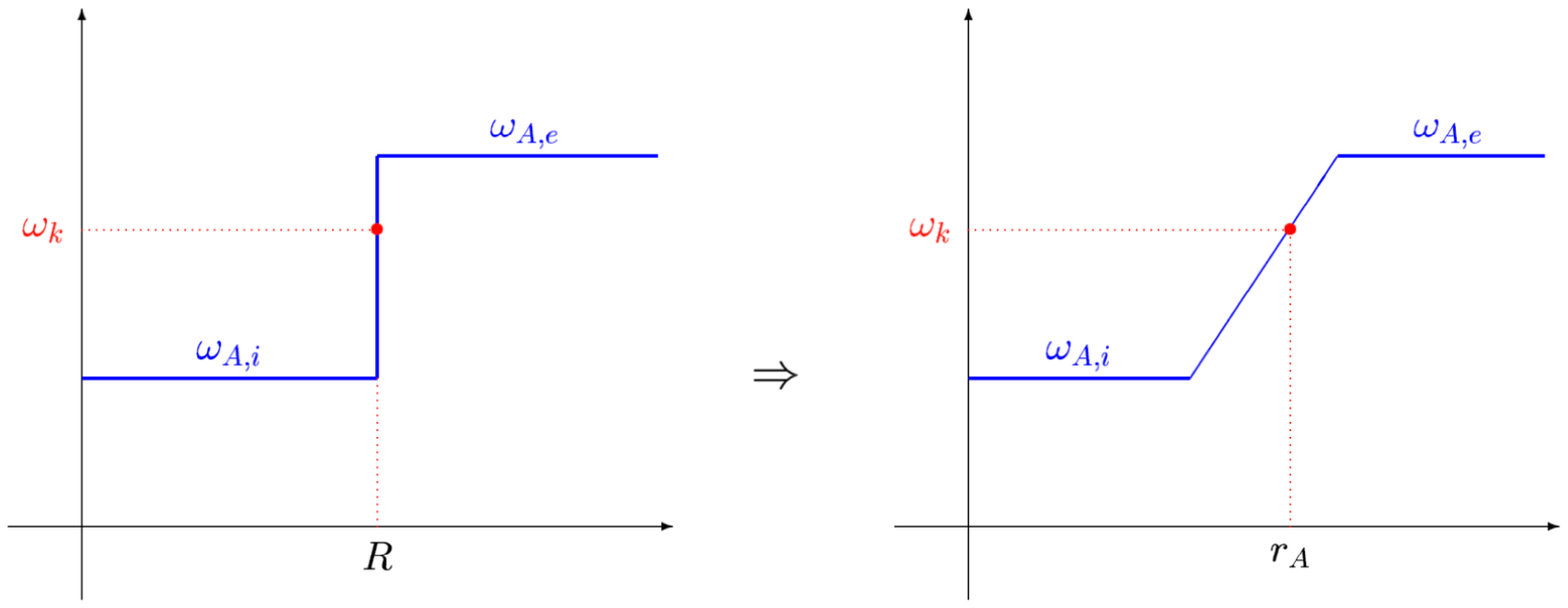}\\
\includegraphics[scale=0.35]{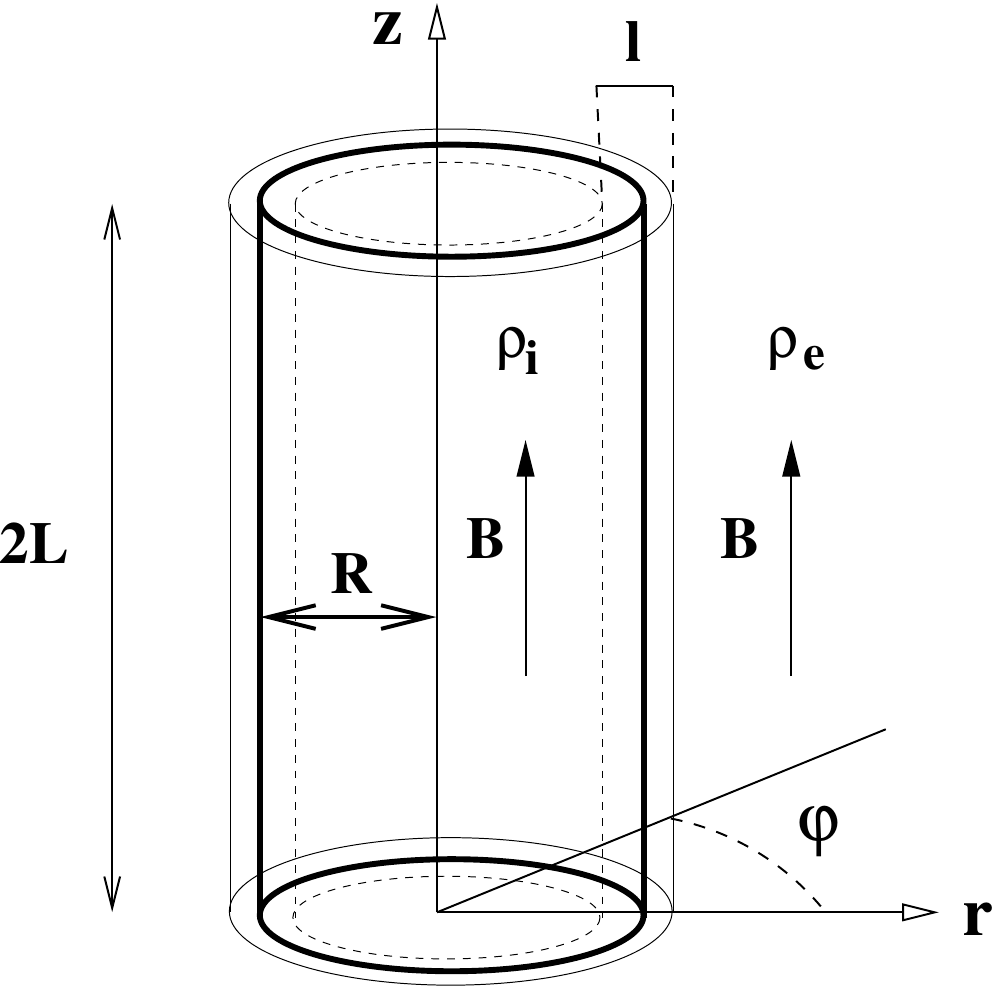}
\hspace{0.2cm}
\includegraphics[scale=0.27]{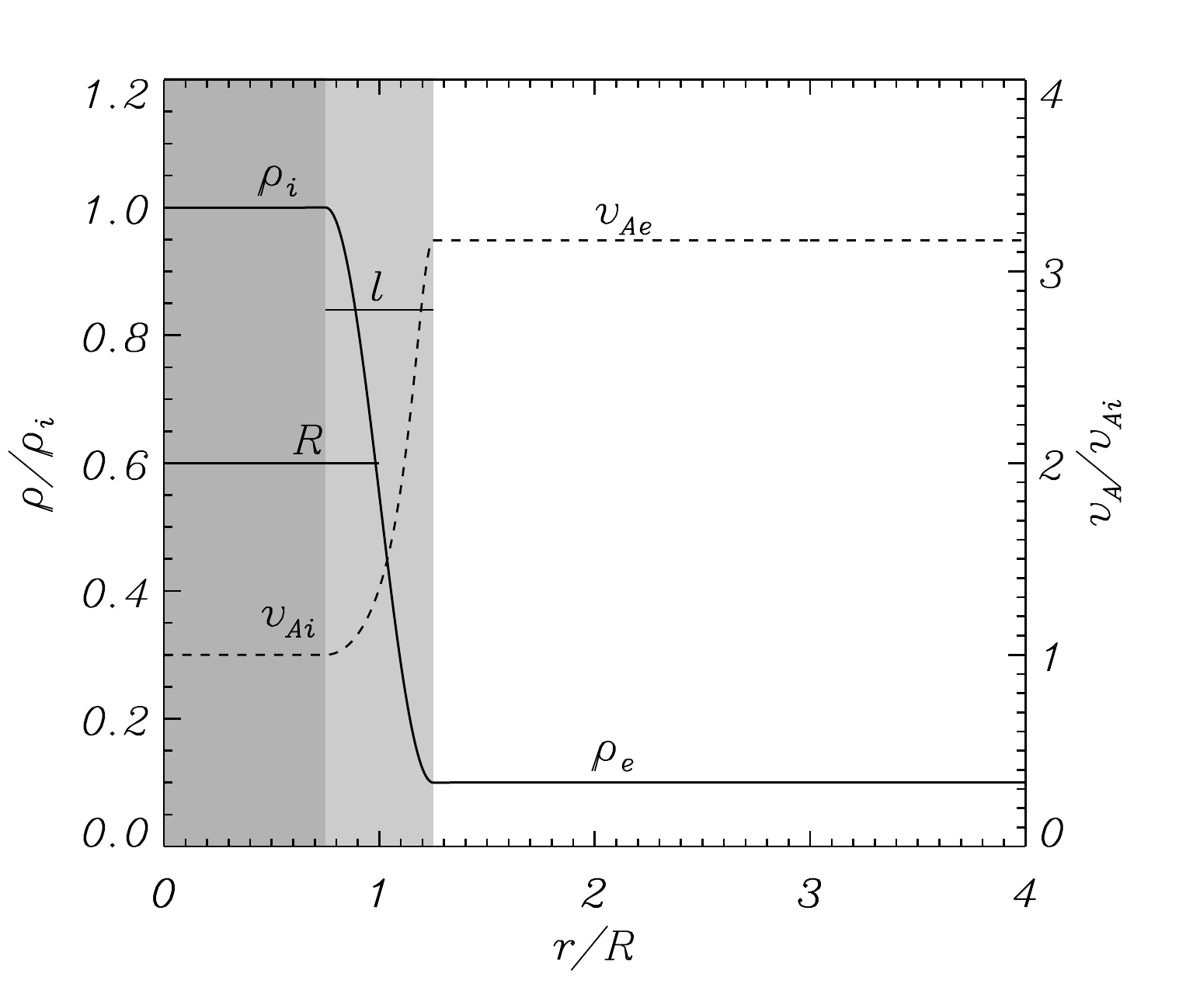}
\caption{Top: schematic representation of the Alfv\'en frequency $\omega_A$ profile in the radial direction in models with a discontinuous jump (left) and with a continuous variation (right) of physical properties. The non-uniformity of the medium makes possible the resonance of the kink mode with the Alfv\'en continuum. Bottom: sketch (left) and radial dependence of physical conditions (right) in the classical cylindrical model representing a straightened coronal flux tube of length 2L and mean radius R modelled as a density enhancement. The magnetic field is uniform and parallel to the $z$-axis and the whole configuration is invariant in the $\varphi$-direction. The density (continuous line) and the Alfv\'en speed (dashed line) vary in a non-uniform boundary layer (light-shaded region) of length $l$ from their constant internal values, $\rho_i$  and $v_{Ai}$ to their constant external values,  $\rho_e$ and $v_{Ae}$ in a nonuniform layer of thickness $l$ defined in the interval [$R-l/2$, $R+l/2$]. Note hat $\rho$ and $v_A$ are normalized to their internal values. In this model $\rho_i/\rho_e=3$, $k_z=\pi/50$ and $l=0.5 R$.}
\label{fig:f1}
\end{figure}

\begin{figure}
\includegraphics[scale=0.8]{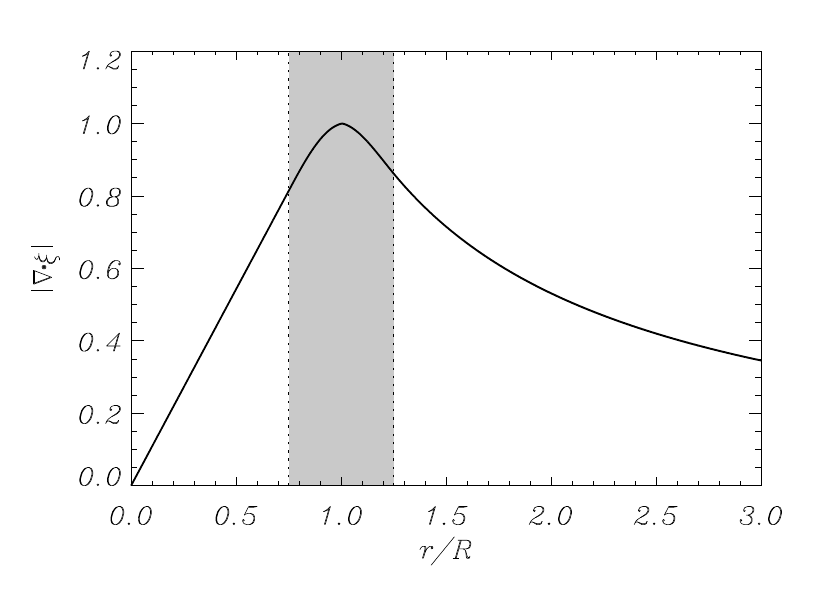}
\caption{Absolute value of the compression as a function of the radial position in a flux tube with $l/R= 0.5$, $k_{\rm z} R = 0.1$, and  $\rho_{\rm i}/\rho_{\rm e} = 3$.  The normalisation $\mbox{max}\left\{\mid \nabla \cdot \xi \mid \right \} = 1$ has been used. The shaded zone denotes the non-uniform region. Credit: \cite{goossens20}, reproduced with permission  © ESO.}
\label{fig:f2}
\end{figure}

\begin{figure}
\includegraphics[scale=0.4]{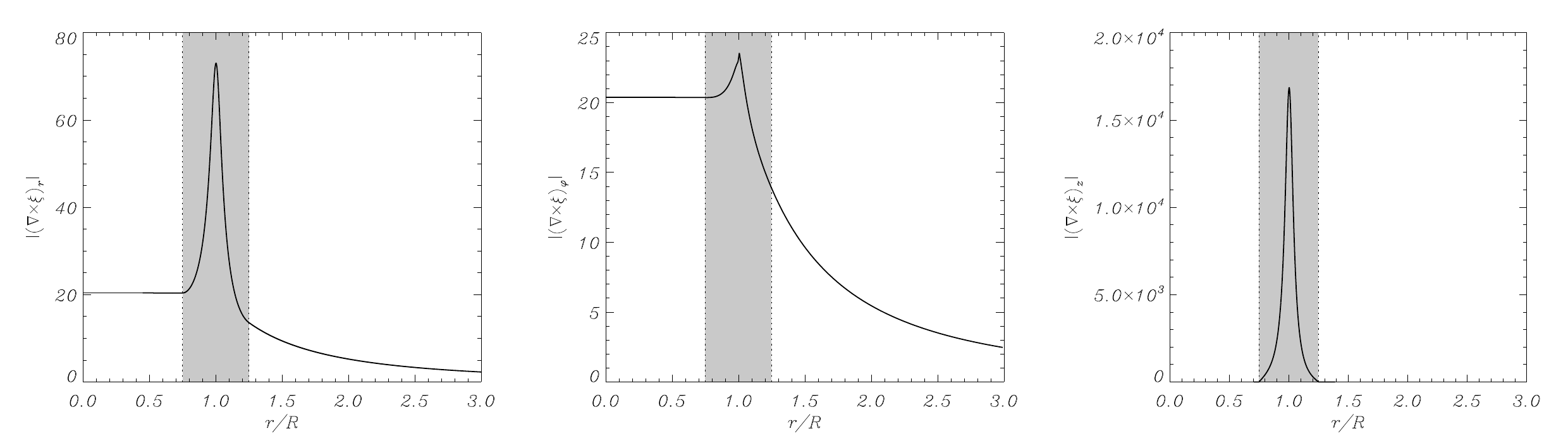}
\caption{Absolute value of the radial (left), azimuthal (centre), and parallel (right) vorticity components as functions of the radial position in the same flux tube as in Figure~\ref{fig:f2}. Credit: \cite{goossens20}, reproduced with permission  © ESO.}
\label{fig:f3}
\end{figure}

\begin{figure}
\includegraphics[scale=0.4]{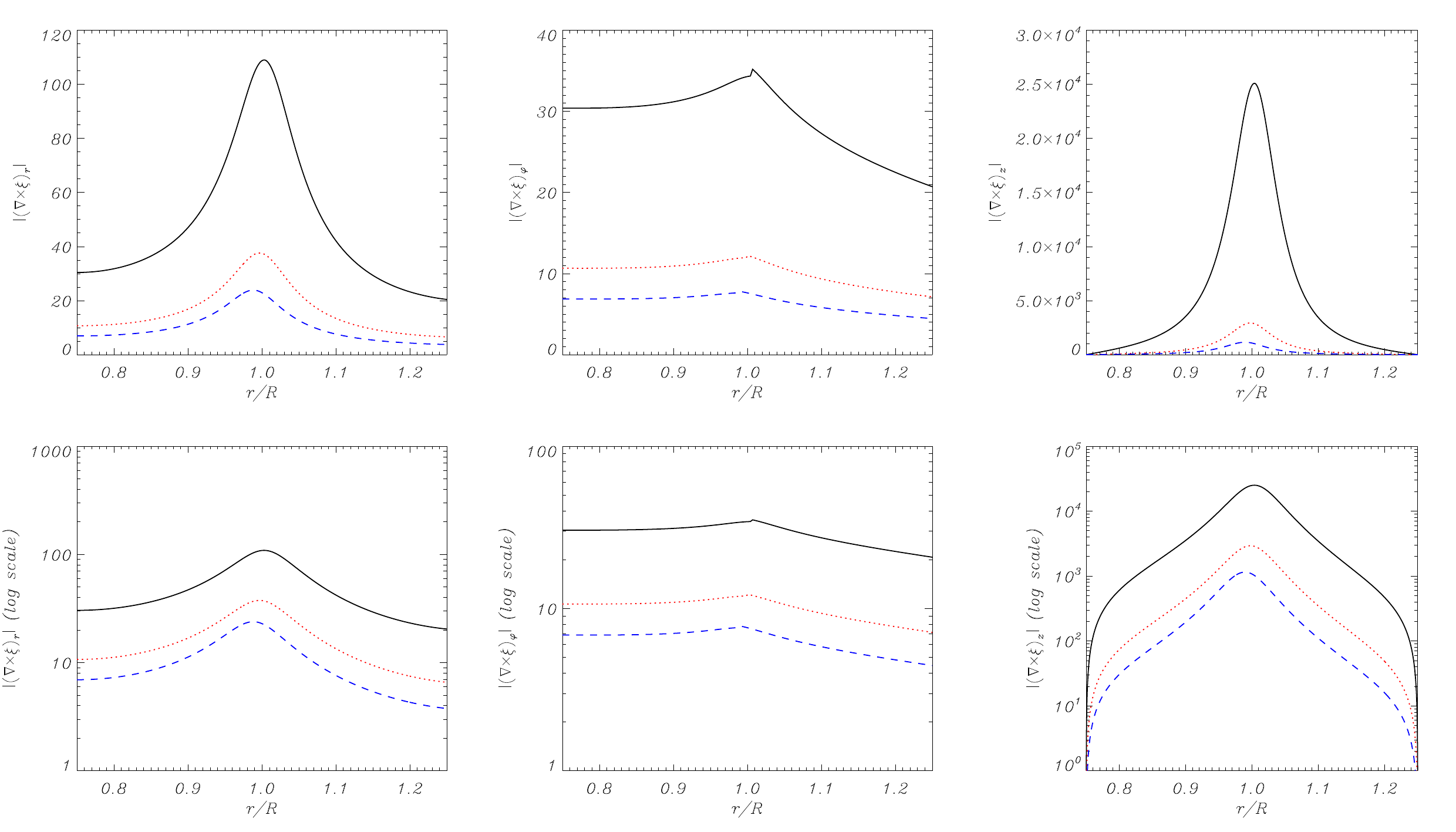}
\caption{Absolute value of the radial (left), azimuthal (centre), and parallel (right) vorticity components in the non-uniform part of a flux tube with $l/R=0.5$ and $\rho_{\rm i}/\rho_{\rm e} = 2$. The top panels are in linear scale, and the bottom panels are in logarithmic scale. The different line styles denote $k_{\rm z}R = 0.1$ (solid black line), $k_{\rm z}R=0.3$ (dotted red line), and $k_{\rm z}R=0.5$ (dashed blue line). Credit: \cite{goossens20}, reproduced with permission © ESO.}
\label{fig:f4}
\end{figure}

\begin{figure}
\includegraphics[scale=0.4]{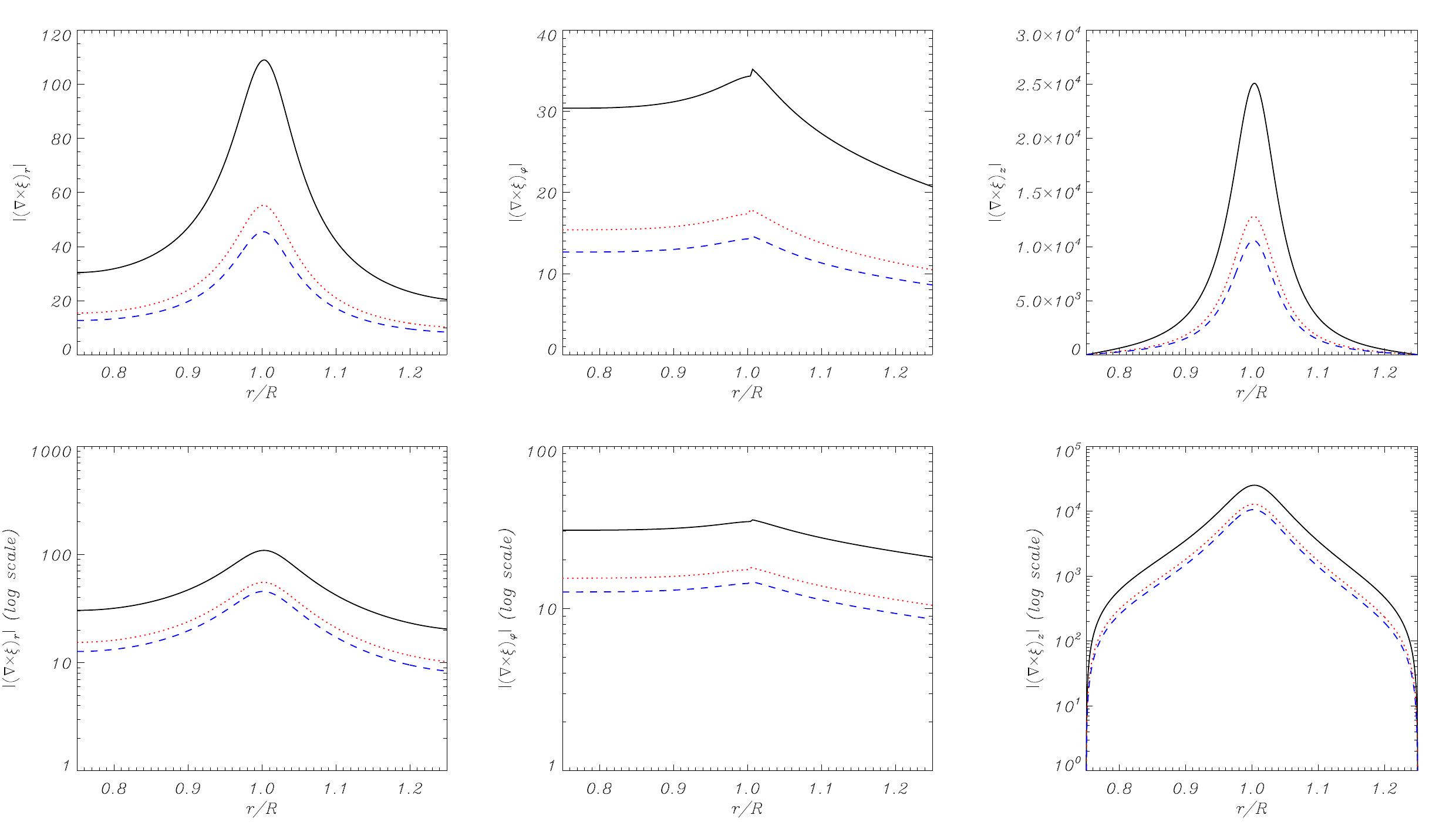}
\caption{Same as Figure~\ref{fig:f4}, but with $k_{\rm z}R = 0.1$ and different values of the density contrast: $\rho_{\rm i}/\rho_{\rm e}=2$ (solid black line), $\rho_{\rm i}/\rho_{\rm e}=5$ (dotted red line), and $\rho_{\rm i}/\rho_{\rm e}=10$ (dashed blue line). Credit: \cite{goossens20}, reproduced with permission © ESO.}
\label{fig:f5}
\end{figure}

\end{document}